\documentclass[12pt]{iopart}
\usepackage{iopams}
\usepackage{iopams,cite}
\usepackage{eurosym}
\usepackage{lipsum}
\usepackage{graphicx}
\usepackage{color}
\usepackage[normalem]{ulem}
\usepackage{lipsum}
\usepackage{tikz}
\usepackage{multicol}
\usepackage{float}
\usepackage{amssymb}
\usepackage{epsfig,color}

\begin{document}
\title[Effects of quantum fluctuations on macroscopic quantum tunnelling ...]{Effects of quantum fluctuations on macroscopic quantum tunneling and self- trapping of BEC in a double well trap}

\author{Fatkhulla  Kh. Abdullaev$^{1}$, Ravil M. Galimzyanov$^{1}$
and Akbar M. Shermakhmatov$^{1,2}$}

\address{
$^1$Physical-Technical Institute of the Uzbekistan Academy of Sciences, Ch. Aytmatov str.,
2-B, 100084, Tashkent-84,  Uzbekistan.\\
$^2$"TIIAME" National Research University,  Kari Niyazov 39, 100000, Tashkent, Uzbekistan
}

\begin{abstract}
   We study the influence of quantum fluctuations  on the
macroscopic quantum tunneling  and
self-trapping of a two-component Bose-Einstein condensate  in a double-well trap.
Quantum fluctuations are described by the Lee-Huang-Yang
term in the modified Gross-Pitaevskii equation. Employing the modified Gross-Pitaevskii equation in scalar
approximation, we derive the dimer model using a two-mode
approximation. The frequencies of Josephson oscillations and
self-trapping conditions under quantum fluctuations are found analytically and
proven by numerical simulations of the modified Gross-Pitaevskii equation.
The tunneling and localization phenomena are  investigated also for
the case of the Lee-Huang-Yang fluid loaded in the double-well potential.
\end{abstract}
\date{\today}

\maketitle

\section{Introduction}
\label{intro}

   Recently much attention has been paid to the study of the beyond
mean-field effects. The contribution to the energy of a BEC due to
quantum fluctuations around the Bogolyubov ground state has been found by
Lee, Huang and Yang (LHY)~\cite{LHY}. While these corrections are
small, in many cases they lead to nontrivial effects. One of
important is their role in the collapsing condensate. It has been
shown, that in the case of two-component condensate, quantum
fluctuations(QF) can stabilize collapsing regime and lead to the
appearance of quantum droplets(QD)~\cite{Petrov, PA}. Generation
of the quantum droplets in the Bose mixture is possible due to the joint
action of the attractive residual mean-field interaction and the
effective repulsion introduced by QF. Another system is a dipolar BEC, where
quantum droplets are generated by the balance between the small
residual mean-field attraction of a combination of local two-body interactions,
 repulsive dipolar interactions and repulsive LHY term~\cite{Dip1}.
 The existence of QD in other systems as the
Bose-Fermi mixtures and Rabi-coupled BEC is discussed in works~\cite{BF, Rabi}.

The influence of the beyond mean-field (BMF) contribution has been
studied for the collective excitations spectrum in Ref.~\cite{collexc},while
modulational instability of matter waves has been studied in Ref.~\cite{MI}.

In this regard, it is interesting to study the influence of BMF
contribution to another phenomena in the BEC. An important case is the processes of macroscopic quantum tunneling
(MQT) and self-trapping(ST) in BEC loaded in a double-well trap
potential. In this system, the nonlinearity due to the interaction
between atoms plays an important role. In particular, it induces
the transition from the Josephson oscillations(JO) of atomic
imbalance to the nonlinear self-trapping (ST) regime~\cite{TM}.
The quantum fluctuations induce an additional nonlinear interaction in the form of
effective
repulsion. Thus in some region of parameters, we can expect the
processes of tunneling in the double-well potential to be very sensitive to the
action of quantum fluctuations. In this problem, we have a few small
parameters: the tunneling parameter(describing the amplitude of tunneling between condensates),
the strength of QF given by the LHY correction, and the residual mean field nonlinearity between
components of BEC. When they are comparable, we can expect new regimes in the switching of
the matter waves in the double-well potential. The important
limiting case of the vanishing of residual mean-field interactions
 is a LHY fluid~\cite{Jor} loaded in the
double-well potential.  Note that recently the LHY fluid has been observed in Ref.
~\cite{Skov}

In this work, we  study the MQT and ST processes in a two-component quasi-one-dimensional  BEC
loaded into a double-well trap, when  quantum fluctuations are
taken into account.

  The structure of the paper is as follows.

In Section~\ref{sec:aver}, we describe the modified
quasi-one-dimensional GP equation, obtained for two-component BEC
in a scalar approximation and including the LHY term. In Section~\ref{sec:twomodes}, two-mode(dimer) model for BEC loaded into the
double-well trap is derived. The Hamiltonian, fixed
points, the Josephson oscillations  are analyzed in Section~\ref{dimer}. The nonlinear self-trapped regime is discussed in
Section~\ref{sec:Selftrap}. In Section~\ref{sec:results}, results of
 full numerical simulations of the modified Gross-Pitaevskii(GP) equation with the
double-well potential are presented.
The conditions for experimental observation of the investigated effects are discussed.

\section{The beyond mean field model for a BEC in a double-well  potential}
\label{sec:aver}

We  consider the dynamics of a two-component BEC loaded in a
cigar-type trap, when the beyond mean-field effects, describing
quantum fluctuations, are taken into account. Assuming that the
wave functions are related by:
$$
\Psi_1=\sqrt{\frac{g_{11}}{g_{22}}}\Psi_2=\Psi,
$$
where $g_{ij}=4\pi \hbar^2 a_{ij}/m$, $a_{ij}$ are inter- and
intra-species atomic scattering lengths, and supposing
$g_{11}=g_{22}=g$ and  $g_{12}=\delta g +g$, we  reduce the coupled
GP equations to the scalar GP equation for $\Psi$. Then the
governing equation, describing the wave function of a condensate
with the  Lee-Huang-Yang term, is~\cite{Petrov}:
\begin{equation}
i\hbar\Psi_t(r,t) = -\frac{\hbar^2}{2m}\nabla^2\Psi(r,t) +
V(r)\Psi(r,t) -\delta g|\Psi|^2\Psi + \gamma_{LHY}|\Psi|^3\Psi,
\end{equation}
where
$$
 \gamma_{LHY}=\frac{128\sqrt{\pi}\hbar^2}{3m}|a|^{5/2}.
$$
We  study below a quasi-one-dimensional geometry given by a
cigar-type trap, when the transversal ($\omega_\perp$) and longitudinal ($\omega_x$) trap frequencies are satisfied to the
condition $\omega_{\perp}\gg \omega_x$. Here we  use the
factorization~\cite{Michinel}:
\begin{equation}
\Psi=\Phi(x,t)R(\rho), \ R(\rho)=Ae^{-\rho^2/2l_{\perp}^2}, \
l_{\perp}=\sqrt{\hbar/(m\omega_{\perp})}. \
\end{equation}
After integration over the transverse distribution, we obtain the
quasi-one-dimensional MGP equation:
\begin{equation} \label{MGPE}
i\hbar\Psi_t(x,t) =
-\frac{\hbar^2}{2m}\Psi_{xx}(x,t)+V(x)_{ext}\Psi(x,t)
-\frac{\delta g}{2\pi l_{\perp}^2} |\Psi|^2\Psi
+\frac{2\gamma_{LHY}}{5\pi^{3/2}l_{\perp}^3}|\Psi|^3\Psi,
\end{equation}
where $V_{ext}(x)$ is a double well  potential chosen as
\begin{equation}\label{VdW}
V_{ext}(x)=\frac{mw_x^2}{2}x^2+ U_0 \exp(-(x/l_0)^2).
\end{equation}

It is useful to introduce dimensionless variables according to:
$$
t=\omega_x t, x=\frac{x}{l_x}, \
l_x=\sqrt{\frac{\hbar}{m\omega_x}}, \psi =
\sqrt{\frac{l_x}{N}}\psi.
$$

Then we can rewrite the governing equation as follows:
\begin{equation} \label{eq1D}
i\psi_t=-\frac{1}{2}\psi_{xx}+V_{ext}(x)\psi + \delta g
|\psi|^2\psi + \gamma |\psi|^3\psi ,
\end{equation}
where
\begin{eqnarray*}
\delta g = \frac{4 \delta a}{l_x} \frac{\omega_{\perp}}{\omega_x} N, \gamma = \frac{128}{15\pi}(\frac{a}{l_x})^{5/2} (\frac{\omega_{\perp}}{\omega_x})^{3/2} N^{3/2},\\
V_{ext}=\frac{1}{2}x^2 + V_0 e^{-(x/l)^2}, l=\frac{l_0}{l_x},
V_0=\frac{U_0}{\hbar\omega_x}.
\end{eqnarray*}
This normalization of the wave function $\psi$ corresponds to:
$$
\int_{-\infty}^{\infty}|\psi|^2 dx =1.
$$

Let us estimate the parameters for the experiments, as performed in
Ref.\cite{Skov}. The atomic scattering length is about $100a_0$, where
$a_0$ is the Bohr radius, $l_x \approx 300\mu m$, $N=10^2 \div
10^3$. The values of parameters $g=0 \div 10, \ \gamma = 0.01 \div
1$.

\section{Two-mode model}
\label{sec:twomodes} To describe the dynamics of the BEC we will use
the two-mode model~\cite{TM}. Let us consider the solution in the
form:
\begin{equation} \label{eq2mode}
    \psi =u(t)\Phi_1(x) +v(t)\Phi_2(x),
\end{equation}
where $\Phi_{1,2}(x)$ are determined by the solutions of the
stationary GP equation. Firstly we obtain two solutions of the
stationary equation, (ground state with the norm $N_{gr}=1$ and an
excited state with the same norm $N_{exc}=1$)

\begin{eqnarray}\label{eqST}
[-\frac{1}{2}\frac{d^2}{dx^2} + V_{ext}(x) +  \delta g |\Phi_{gr/exc}|^2 + \nonumber\\
\gamma |\Phi_{gr/exc}|^3 -  \mu ]\Phi_{gr/exc}=0.
\end{eqnarray}
It should be noted that a ground state function and an excited one
are chosen to be symmetric and antisymmetric, respectively,
$\Phi_{gr}(-x)=\Phi_{gr}(x)$ and $\Phi_{exc}(-x)=-\Phi_{exc}(x)$.
   Then, functions $\Phi_{1,2} (x)$, localized in corresponding wells
(1,2) of the double-well potential are expressed as
\begin{eqnarray}\label{Phi12}
\Phi_1 (x) = (\Phi_{gr}(x)  + \Phi_{exc}(x))/\sqrt{2},\\
\Phi_2 (x) = (\Phi_{gr}(x) - \Phi_{exc}(x))/\sqrt{2}, \\
I = \int_{-\infty}^{\infty} \Phi_i(x) \Phi_j (x)dx, \ i,j=1,2.
\end{eqnarray}
where $I$ is the overlap integral.

\begin{figure}[htbp]
\centering
\includegraphics[width=6.0cm,angle=-90,clip]{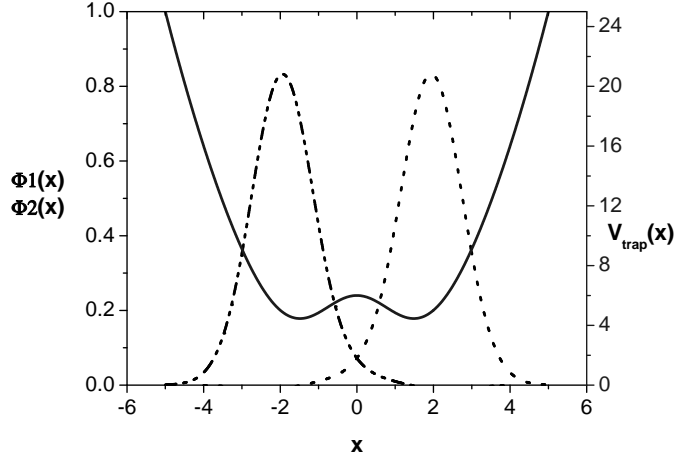}
\caption{The double-well potential with parameters $V_0=5, \
l=1.5$. The solid line is for the double-well potential, -
dash-dot-dot and dot- lines are for $\Phi_2(x)$ and $\Phi_1(x)$
wave functions, respectively. } \label{fig:mu}
\end{figure}

In Fig.~\ref{fig:mu} the double-well potential with wave functions
$\Phi_{1,2}(x)$ are depicted. Note that the two-mode approximation
works well for the small mean field nonlinearity and overlaps~\cite{AB06, Bao}

Let us substitute the solution ~\ref{eq2mode} into Eq.~(\ref{MGPE}),
and multiply by $\Phi_1$, and integrate over $x$ in the interval
$(-\infty,+\infty)$. Supposing a weak overlap criterion and
ignoring nonlinear overlap integrals, we obtain the following
equation:
\begin{equation}\label{equ}
i u_t=E_1^0u-Kv-\alpha|u|^2u+ \beta |u|^3 u.
\end{equation}
In a similar way, the equation for $v(t)$ can be obtained
\begin{equation} \label{eqv}
i v_t=E_2^0v-Ku-\alpha|v|^2v+ \beta |v|^3 v.
\end{equation}
Here $E_{1,2}^0$ are the zero-point energies in each well, for the
symmetric trap $E_1^0=E_2^0$. K is the amplitude of the tunneling,
$\alpha$ and $\beta$ are other nonlinearity parameters
    \begin{equation} \label{E01}
        E_1^0=\frac{1}{2}\int_{-\infty}^{\infty} |\Phi_{1,x}|^2 dx + \int_{-\infty}^{\infty} \Phi_1 V(x )\Phi_1 dx,
        \end{equation}
        \begin{equation} \label{Kt}
        K=-\left( \frac{1}{2}\int_{-\infty}^{\infty} \Phi_{1,x}\Phi_{2,x} dx+\int_{-\infty}^{\infty} \Phi_1 V(x) \Phi_2 dx \right),
        \end{equation}
        \begin{equation} \label{albe}
        \alpha = \delta g \int |\Phi_1|^4 dx, \
        \beta= \gamma \int \Phi_1^5 dx.
        \end{equation}
The condition of the applicability of the two-mode model is $\mu <
V_0$, where $\mu$ is the chemical potential. For more details see the work ~\cite{AB06}, where the different extensions of the two-mode model are considered.

\section{Dimer model with LHY terms}
 We consider below the symmetric dimer case, when $\Delta
E=E_1^0-E_2^0 = 0$. Then eqs.(\ref{equ}),(\ref{eqv}) are written as: \label{dimer}
\begin{eqnarray} \label{dimereq}
iu_t + Kv + \alpha |u|^2u  -\beta |u|^3u-E_1^0u =0,\\
iv_t + Ku +\alpha |v|^2 v -\beta |v|^3 v-E_1^0v=0.
\end{eqnarray}
Introducing  variables:
$$
u=\sqrt{N_1} e^{i\phi_1}, \  v = \sqrt{N_2} e^{i\phi_2},
$$
we obtain coupled equations for the relative imbalance $Z=(N_1-N_2)/N$
and the relative phase $\psi=\phi_2-\phi_1$:
\begin{eqnarray}
Z_t=-2K\sqrt{1-Z^2}\sin{\psi},\\
\psi_t = 2K\frac{Z}{\sqrt{1-Z^2}}\cos{\psi}+\alpha NZ -\frac{\beta
N^{3/2}}{2^{3/2}}[(1+Z)^{3/2}-(1-Z)^{3/2}].
\end{eqnarray}
Introducing  variables:
$$
\Lambda = \frac{\alpha N}{2K}, \ \epsilon = \frac{\beta
N^{3/2}}{2K}, \ t=2K t,
$$
we get:
\begin{eqnarray}
Z_t=-\sqrt{1-Z^2}\sin{\psi}, \label{eqz} \\
\psi_t=\frac{Z}{\sqrt{1-Z^2}}\cos{\psi}+\Lambda Z -
\frac{\epsilon}{2^{3/2}}[(1-Z)^{3/2} -(1+Z)^{3/2}]. \label{eqphi}
\end{eqnarray}

This system has the Hamiltonian form:
\begin{equation}\label{sys1}
Z_t= -\frac{\partial H}{\partial\phi}, \ \psi_t = \frac{\partial
H}{\partial Z},
\end{equation}
with the Hamiltonian
\begin{equation}\label{Hamilt}
H = -\sqrt{1-Z^2}\cos{\psi} +\frac{\Lambda}{2}Z^2 +
\frac{\epsilon}{5\sqrt{2}}a,
\end{equation}
where
$$
a=[(1+Z)^{5/2}+(1-Z)^{5/2}].
$$
The case, when the residual mean field
interaction is equal to zero,$\Lambda =0$, corresponds to the Lee-Huang-Yang
fluid, loaded into a double-well potential \cite{Jor}.

\subsection{Josephson oscillations }
\label{Jo_oc} To analyze different regimes in the dynamics of the
atomic population imbalance and the relative phase, let us find
the stationary states  of the system. Equating
the derivatives on time to zero, we have:
$$
\psi_c=2\pi n, \ z_c=0, n=0,1,2,... \ \mbox{with} \
 E_{+} =-1+\frac{\sqrt{2}}{5}\epsilon,
$$
and
$$
\psi_c=(2n+1)\pi, \ z_c=0, n=0,1,2,... \ \mbox{with} \
 E_{-}=1+\frac{\sqrt{2}}{5} \epsilon,
$$
where $E_{+/-}$ are the values of Hamiltonian
~(\ref{Hamilt}) at the critical points $Z_c, \ \psi_c$.
From Eq.~(\ref{sys1}), we can find the frequencies of Josephson
oscillations $\omega_J$ of the atomic population imbalance $Z$.
Two cases of the relative phase difference can be studied.

i).The case of a zero-phase mode: $\psi(0)=0, <Z>=0$. Then we obtain
for the frequency of small oscillations Z, i.e. the JO frequency $\omega_J$:
\begin{equation} \label{zerophase}
\omega_J =\sqrt{1+\Lambda +\frac{3\sqrt{2}\epsilon}{4}}.
\end{equation}

ii).The case of a $\pi$-phase mode: $\psi(0) =\pi, \ <z>=0.$ The
frequency of the JO is:
\begin{equation} \label{piphase}
    \omega_J = \sqrt{1-\Lambda -\frac{3\sqrt{2}\epsilon}{4}}.
\end{equation}

 If the strength of the residual mean field nonlinearity and the
LHY corrections are satisfied to the relation:
 \begin{equation} \label{quasilinear}
 \Lambda =-\frac{3\sqrt{2}\epsilon}{4}, \ \alpha = \frac{3\beta
N^{1/2}}{2^{3/2}},
 \end{equation}
then the frequency of Josephson oscillations is close to the
frequency of oscillations of the unperturbed system (the Rabi
oscillations case). It corresponds to the quasi-linear regime.
Thus it is open the possibility to measure the strength of quantum fluctuations strength
by properly detuning the residual mean-field scattering length
$\delta a = -|a_{12}| + a.$

If we linearize Eq.~(\ref{sys1}) on $Z$ only,  the next equation
for the phase is obtained:
\begin{equation}
\psi_{tt}+(\Lambda + \frac{3\epsilon}{2^{3/2}})\sin(\psi)
+\frac{1}{2}\sin(2\psi)=0.
\end{equation}
This equation has the form of an equation of motion of the unit mass particle under
the action of the periodic potential. When $\epsilon =0$,  the result obtained in Ref.~\cite{TM} is
reproduced. At $\psi=\pi$($\pi$-phase mode), in the potential a valley exists, where the effective
particle oscillates. By properly detuning the  parameter $\Lambda \rightarrow 1 -3\epsilon/2^{3/2}$,
we can diminish this valley to zero. The nonlinearity breaks the $z$-symmetry
and leads to the appearance of  states with $\langle Z \rangle \neq 0$.

\subsection{Self-trapped regime}
\label{sec:Selftrap} The self-trapped regime corresponds to the
localization of atoms in one of the wells, i.e. to $\langle Z \rangle \neq 0$.
If the population imbalance is fixed, then we can obtain the condition on the nonlinearity
parameter. It can be find from the condition:

\begin{equation}
    H > 1 + \frac{\sqrt{2}\epsilon}{5}.
\end{equation}
Using Eq.~(\ref{Hamilt}), we obtain the condition for self-trapping:
\begin{equation}
\Lambda > \Lambda_c = \frac{2}{Z(0)^2}[1 +
\frac{\epsilon}{5\sqrt{2}}(2-a)+\sqrt{1-Z(0)^2}\cos(\psi(0))].
\end{equation}

For the LHY fluid case, the localization criterion is:
\begin{equation}\label{crit1}
\epsilon_c =
\frac{5\sqrt{2}(1+\sqrt{1-Z(0)^2}\cos(\psi(0)))}{a-2}.
\end{equation}

\begin{figure}[htbp]
\includegraphics[width=6.5cm,clip]{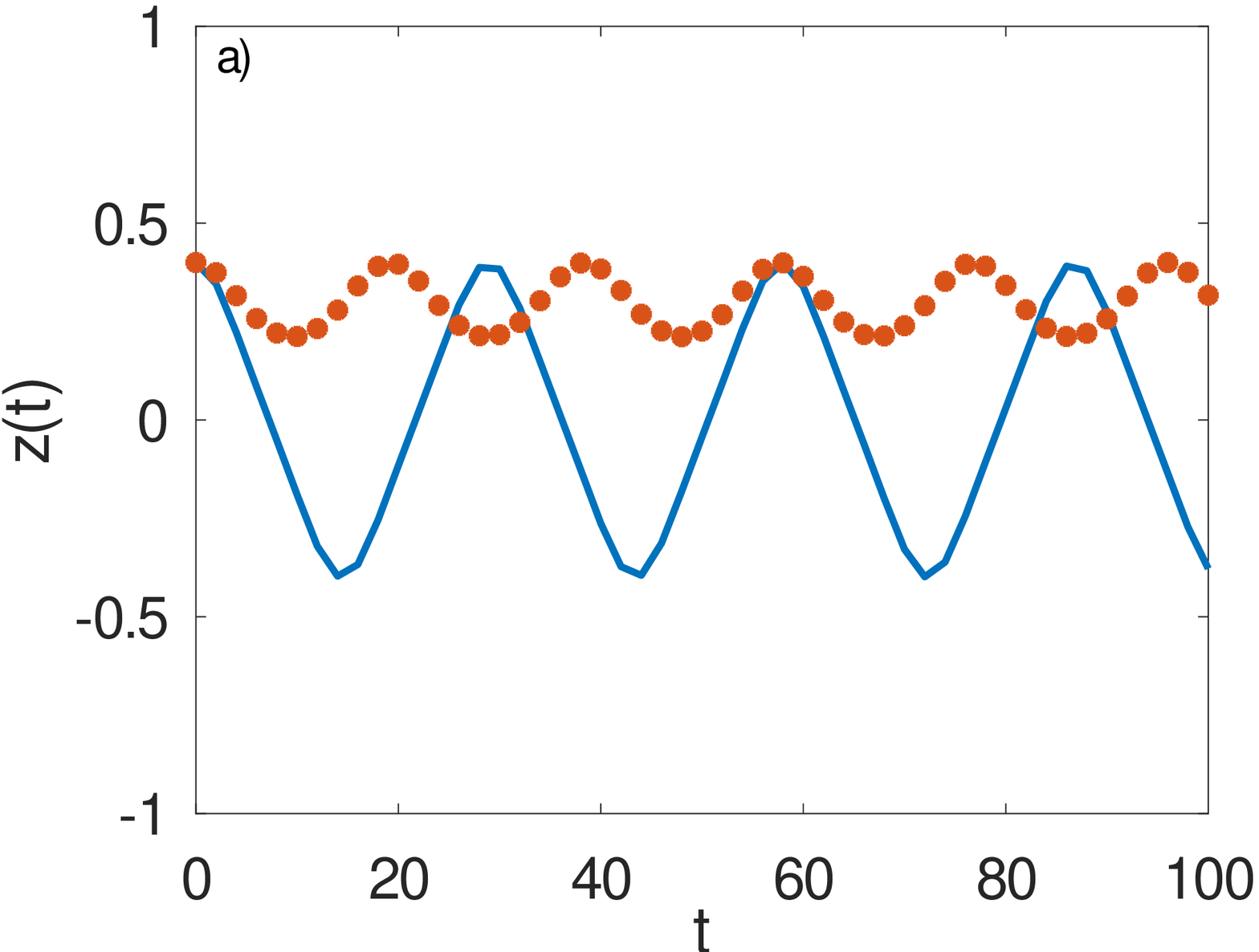}\quad
\includegraphics[width=6.0cm,clip]{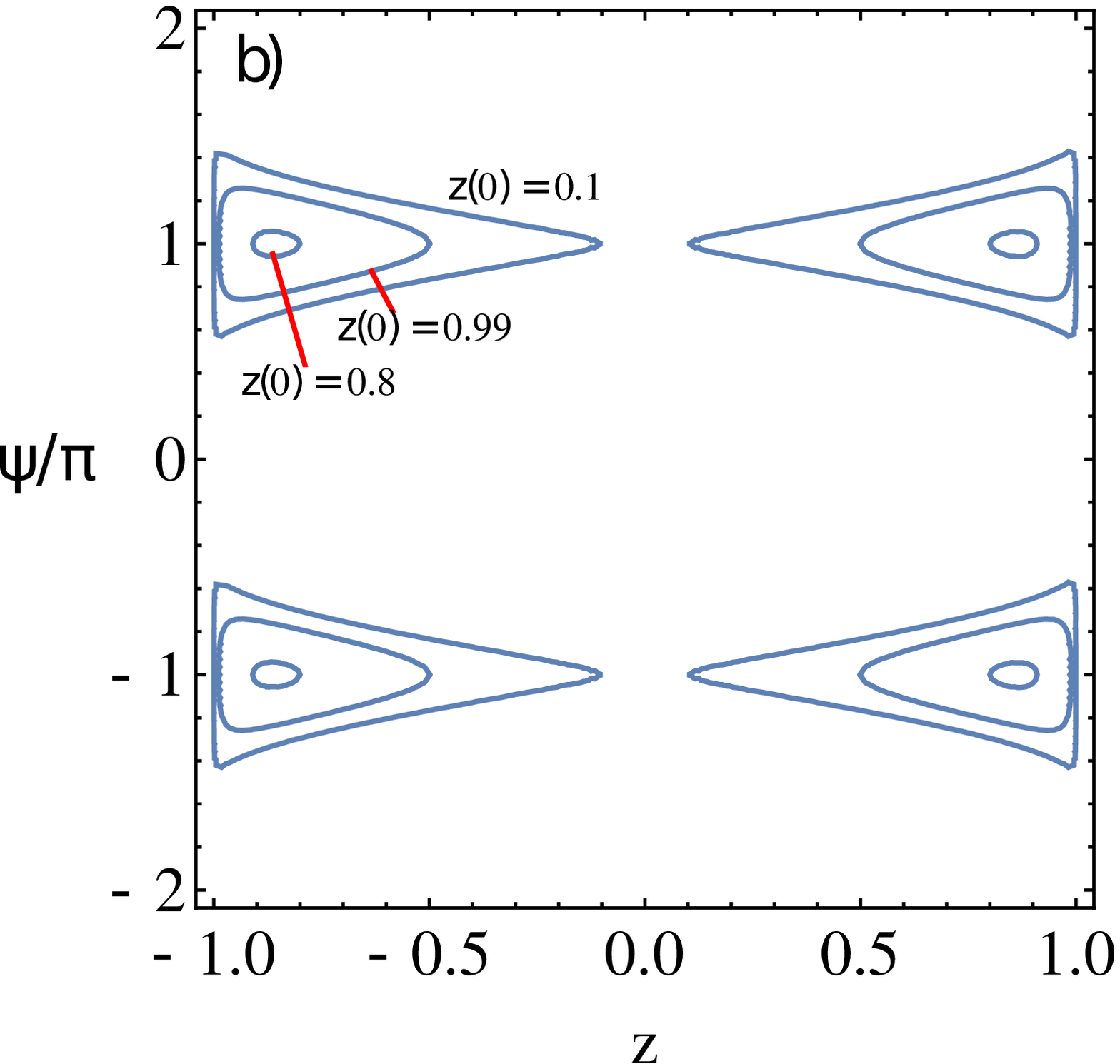}\quad
\caption{ a) The dynamics of the relative imbalance $Z(t)$ for the
LHY fluid ($\delta g =0$). Dot-line is for $\epsilon=1$ and the solid
line is for $\epsilon=0.96$. Critical value $\epsilon_c=0.9872$.
For both cases, oscillations are in the $\pi$-phase regime and the
initial imbalance $Z(0)=0.4$. b) Phase portrait for the LHY fluid,
where $\epsilon=1.96$.} \label{LHYfluid}
\end{figure}

In Fig.~\ref{LHYfluid}a the dynamics of the relative imbalance $Z$
for the LHY fluid ($\Lambda =0$)in $\pi$-phase mode is shown. From Eq.(\ref{crit1}) we obtain $\epsilon_c = 0.9872$.
Here the dotted line is for
the case $\epsilon=1$ and the solid line for $\epsilon=0.96$. One can
see that can see that at $\epsilon>\epsilon_c$, the self-trapping
regime appears.

In Fig.~\ref{LHYfluid}b the phase-portrait in ($Z,\psi$) plane for
the LHY fluid case is presented. Is clearly seen the
self-trapping of condensate in one of the wells under the action of
quantum fluctuations. Note, that at these values of parameters, the
self-trapping exists in the $\pi$-phase mode.

%
\section{Numerical simulations}

\label{sec:results} We  also perform  full numerical
simulations of the modified GP equation with the double-well
potential to the study the macroscopic quantum tunneling and
self-trapping phenomena.

We start from the calculation of the ground state and exited state
solutions with $N_{gr}=N_{ex}=1$ for given $\gamma$ and $\delta
g$. Then using Eq.~(\ref{Phi12}) we construct initial wave
function for governing equation Eq.~(\ref{eq1D}):
\begin{equation}\label{Psi0}
\Psi(x,t=0)=u(0) \Phi_1(x) + v(0) \Phi_2(x),
\end{equation}
where $u(0)^2-v(0)^2=Z(t=0)$,
 By computing the full equation Eq.~(\ref{eq1D}) we get
the dependencies $Z(t),\phi(t)$.

For obtaining the relative imbalance $Z(t)$ and relative phase
$\psi(t)$ one needs to calculate $N_{1,2}$ and $\phi_{1,2}$ as
follows:
\begin{eqnarray}\label{zfi}
N_{1}(t) = \int_{-\infty}^0 |\Psi(x,t)|^2dx, \\
N_{2}(t) = \int_0^{+\infty} |\Psi(x,t)|^2dx, \\
\phi_{1}(t) =\arctan{\frac{\int_{-\infty}^0
\Im(\Psi(x,t))|\Psi(x,t)|^2dx}{\int_{-\infty}^0
\Re(\Psi(x,t))|\Psi(x,t)|^2dx}},\\
\phi_{2}(t) =\arctan{\frac{\int_0^{+\infty}
\Im(\Psi(x,t))|\Psi(x,t)|^2dx}{\int_0^{+\infty}
\Re(\Psi(x,t))|\Psi(x,t)|^2dx}}
\end{eqnarray}
where $\Im$ and  $\Re$  are the imaginary and real parts.
For the following calculations the double-well potential
parameters are $V_0=4,6,7$ with $l=1.5$.

\begin{figure}[htbp]
\centerline{
\includegraphics[width=6.0cm,angle=-90]{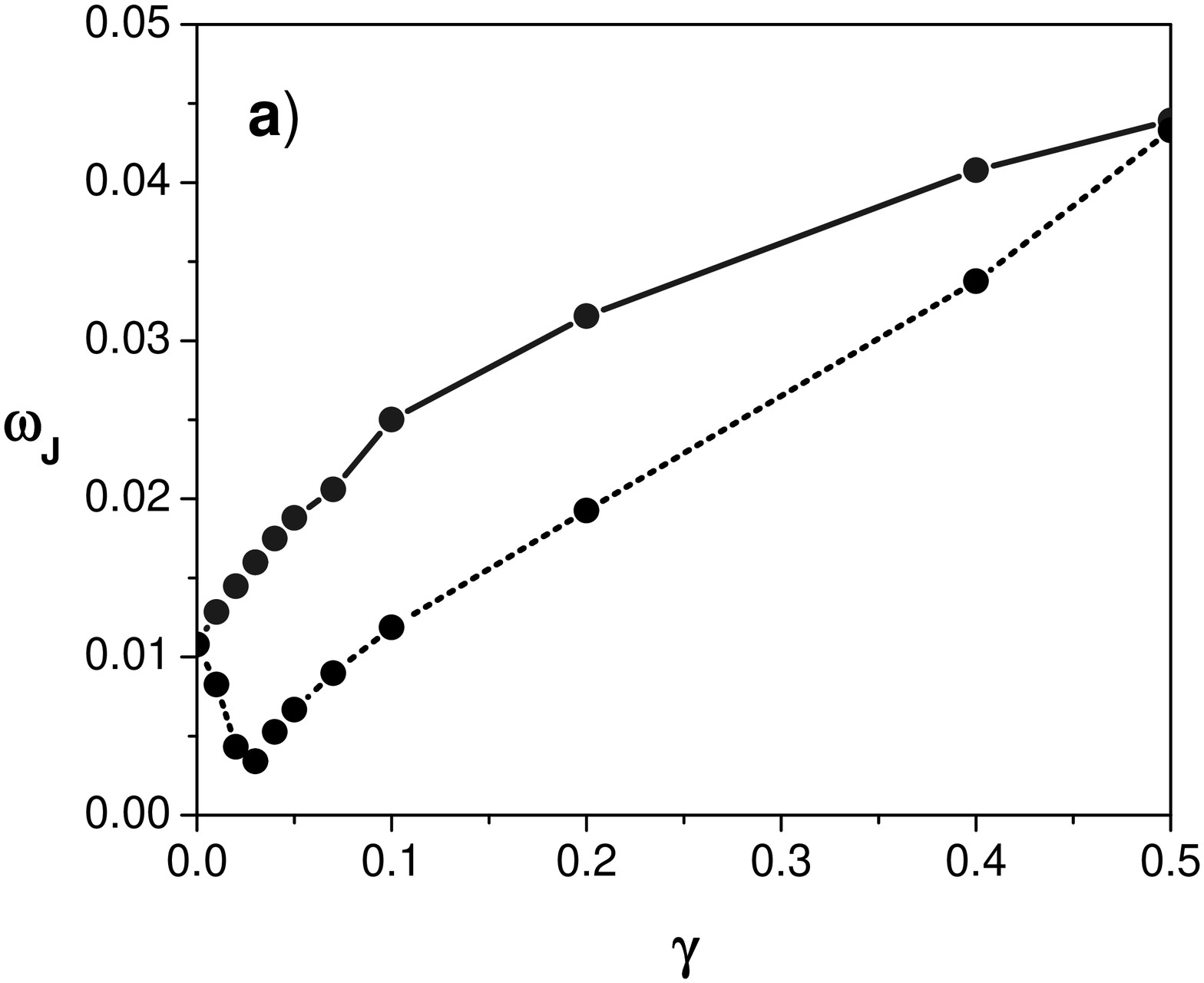}
\includegraphics[width=6.0cm,angle=-90]{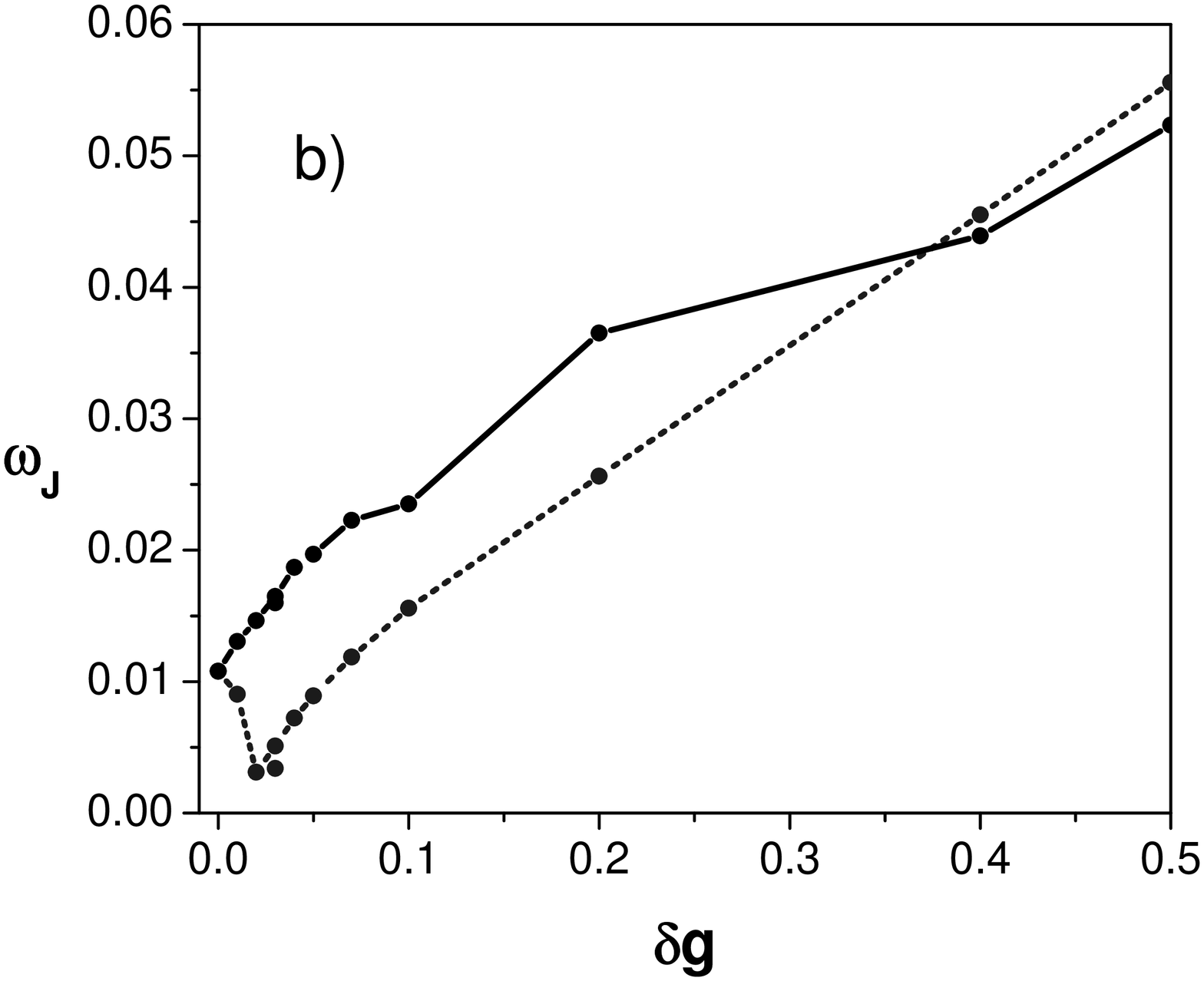} }
\caption{The frequency dependencies of the Josephson oscillations
for a) the LHY-dimer, $\delta g =0$ (left graph) and b) the case
when LHY-term is neglected. In both graphs upper solid lines are
for zero-phase cases and lower dotted lines are for $\pi$-phase
cases. In all cases the initial imbalance of Josephson
oscillations $Z(t=0)=0.02$ for $V_0 = 7, \ l=1.5$.} \label{LHYdg}
\end{figure}

In Fig.~\ref{LHYdg} the dependencies of the JO frequency $\omega_J$
: a) on $\gamma$ (the case of LHY-dimer at $\delta g=0$) and b) on
$\delta g$ (the case when the LHY-term is neglected and $\delta g
\neq 0$) are presented. There are two curves in this figure. The
upper curve is for the frequency dependence of $Z(t)$ oscillations
in the zero-phase regime, when $\psi = 0$, and the lower one is
for the case of $\pi$-phase regime  $\psi=\pi$. One can see that
at the very beginning of the two curves, they are described by
equations Eq.~(\ref{zerophase}) and Eq.~(\ref{piphase})
correspondingly. For both cases $Z(t=0)=0.02$.

As the LHY-term strength $\gamma$ increases, a localization
phenomenon with $\langle Z  \rangle \ne 0$ for $\gamma \approx 0.03$ occurs
breaking the smooth behavior of the lower $\pi$-phase mode curve.

\begin{figure}[htbp]
\centerline{
\includegraphics[width=6.0cm,angle=-90]{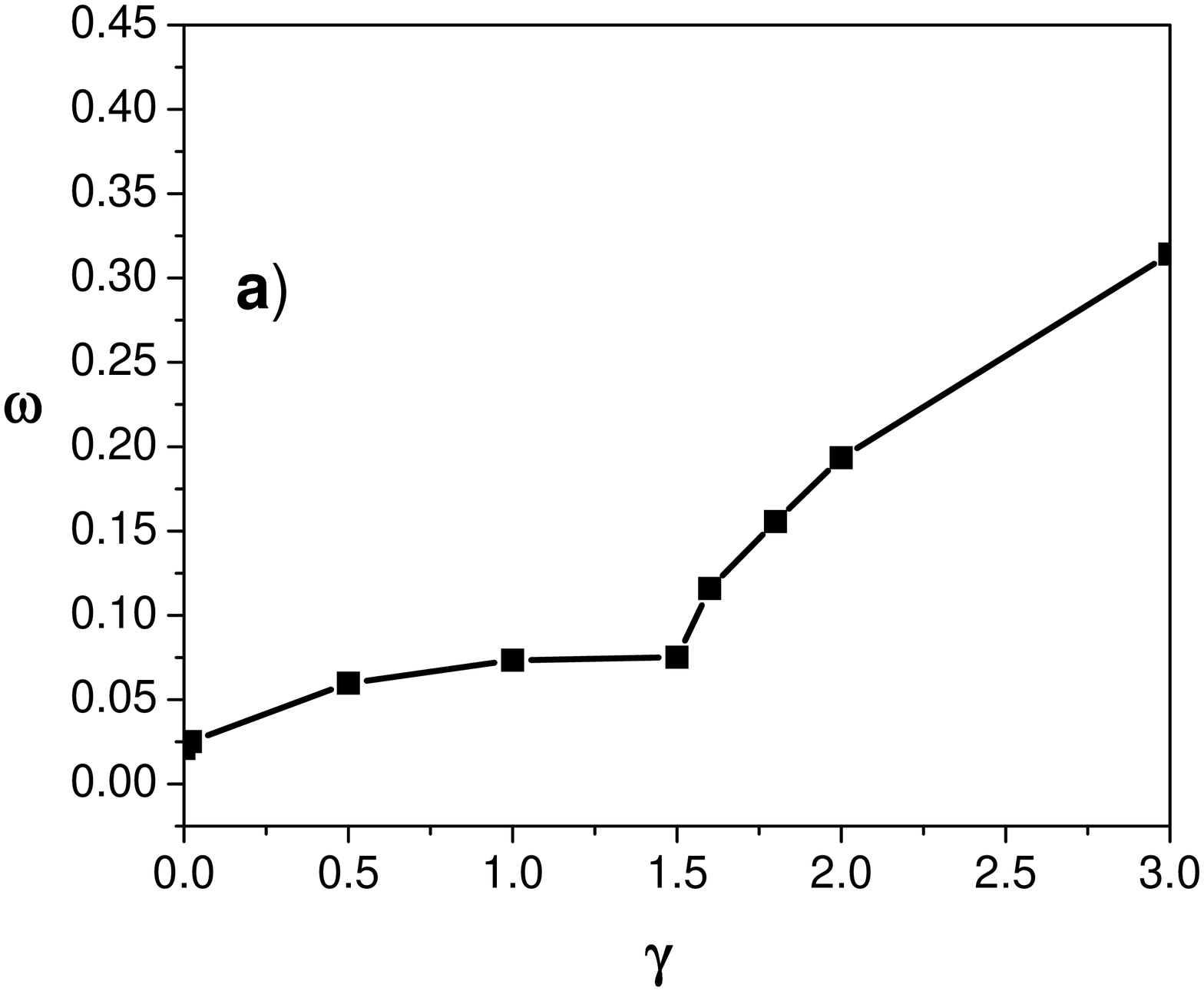}
\includegraphics[width=6.0cm,angle=-90]{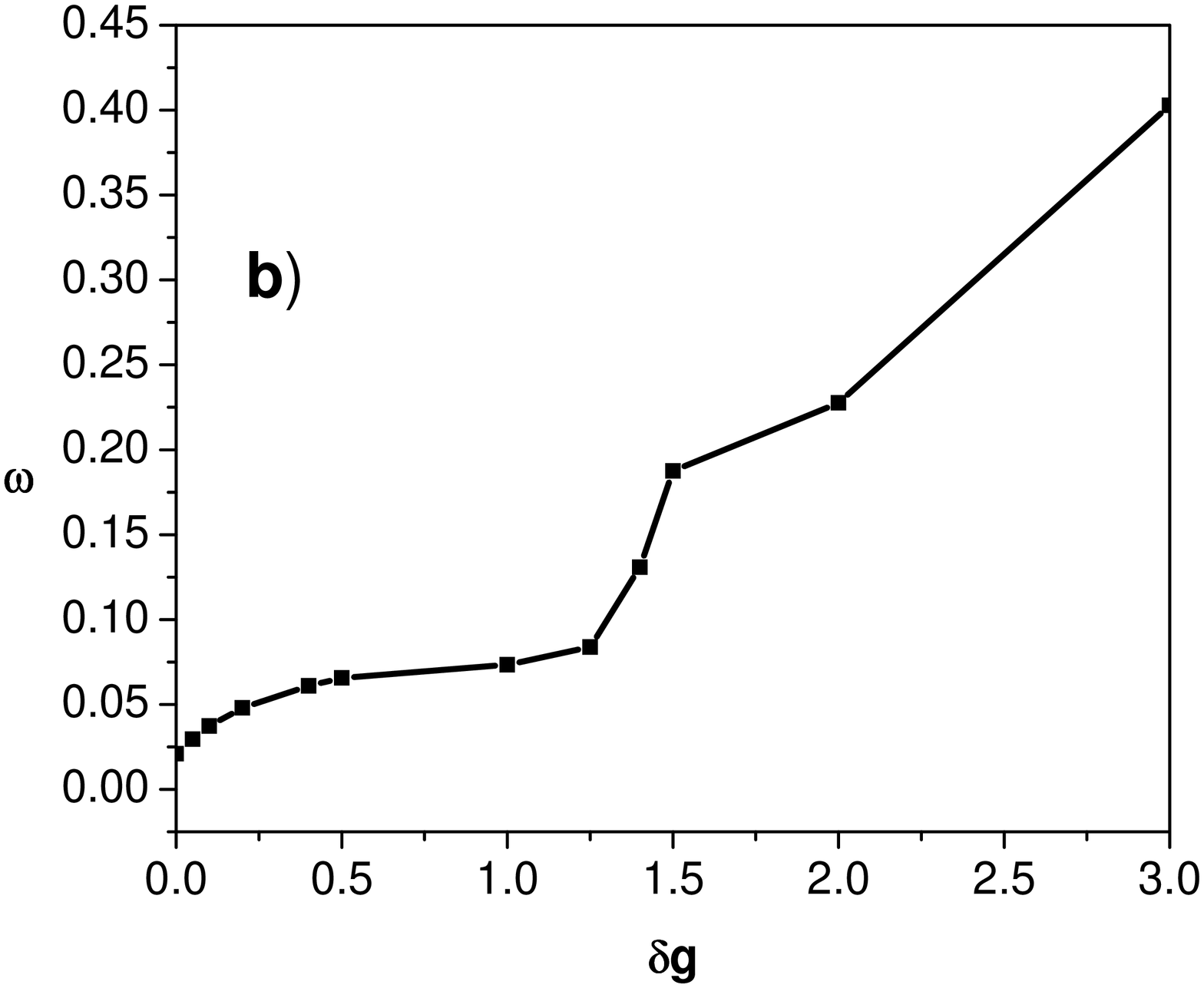}}
\caption{The frequency dependencies of the Josephson oscillations
in the zero-phase mode: a) for different $LHY$-term $\gamma$ when
$\delta g=0$; b) for different nonlinear term $\delta g$ when the
LHY-term is neglected. Parameter $Z(t=0)=0.4$. Double-well
potential parameters $ V_0 = 6, \  l=1.5$ in both cases.}
\label{wLHYdg}
\end{figure}

The frequency dependencies for the same cases, but for larger
intervals of $\gamma$ and $\delta g$ are presented in Fig.~\ref{wLHYdg}.

It should be noted that as seen in the left panel (the LHY fluid
case) for $0 < \gamma < 1.5$ the mean value of the imbalance
$\langle Z \rangle =0$, and $\langle Z \rangle \neq 0$ for $\gamma > 1.6$. In the right panel
(when the LHY term is neglected) for $0 < \delta g < 1.5$, the mean
value of the imbalance $\langle Z \rangle =0$, and $\langle Z \rangle \neq 0$ for $\delta g >
1.5$.

\begin{figure}[htbp]
\includegraphics[width=4cm,angle=-90,clip]{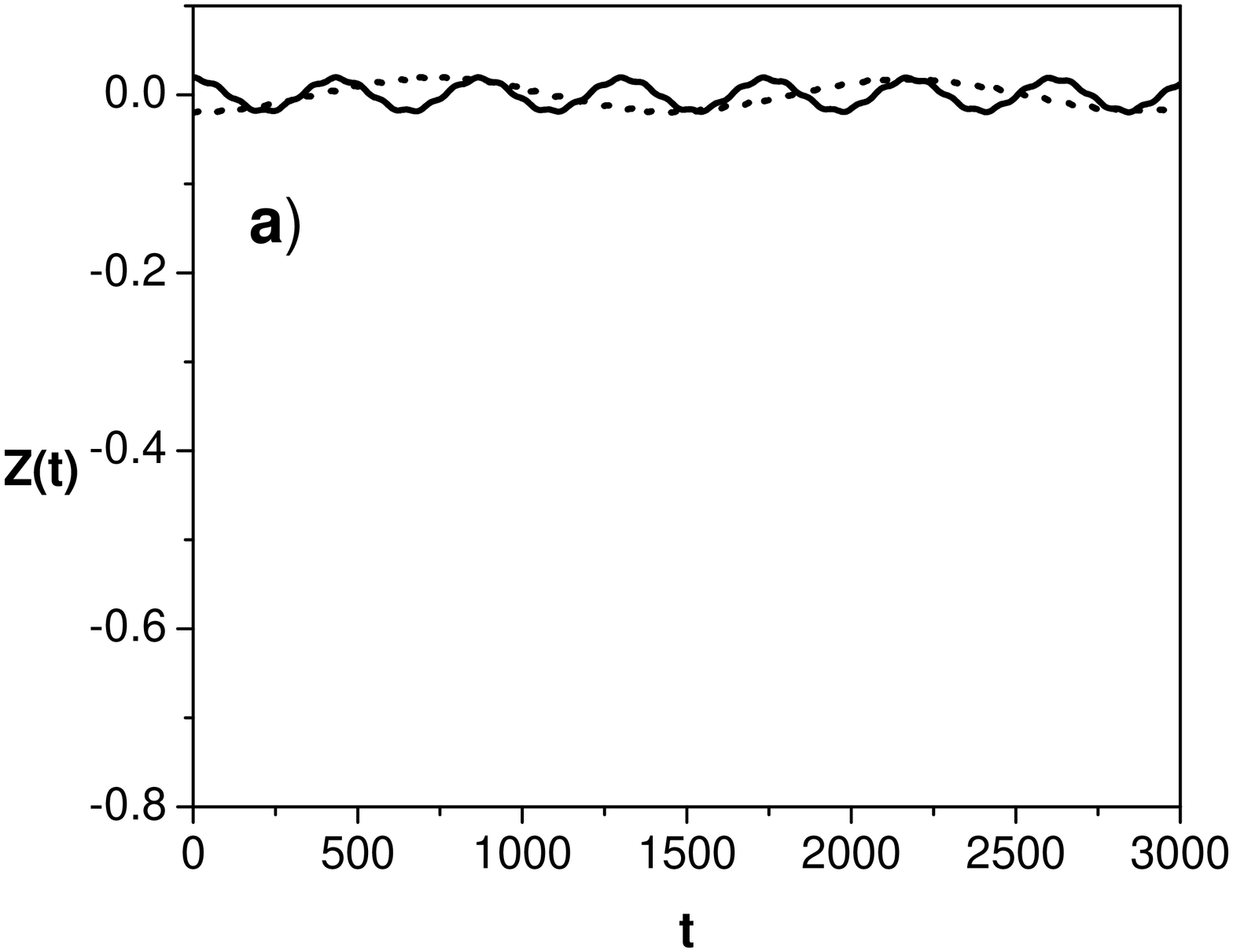}\quad
\includegraphics[width=4cm,angle=-90,clip]{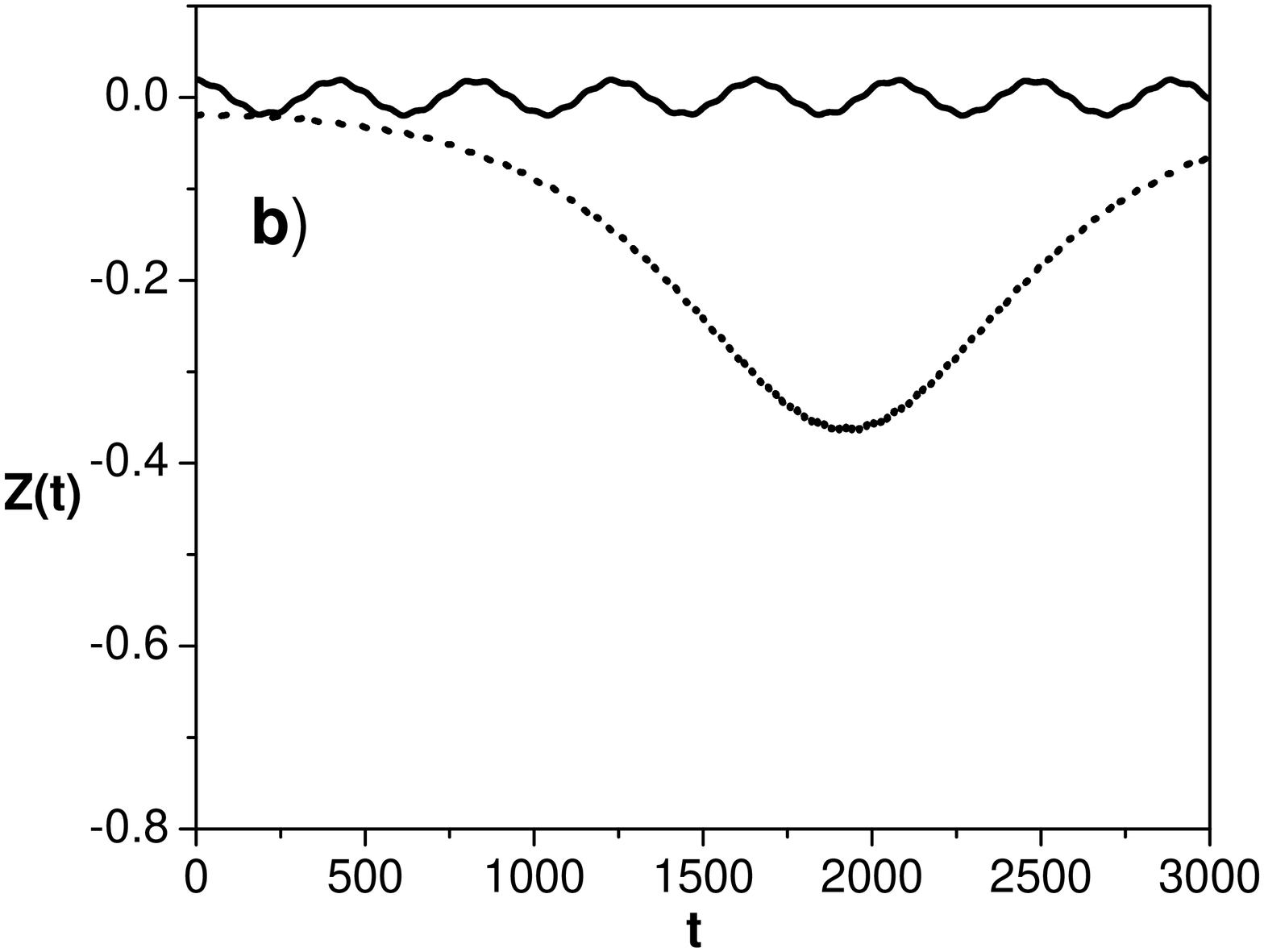}\quad
\includegraphics[width=4cm,angle=-90,clip]{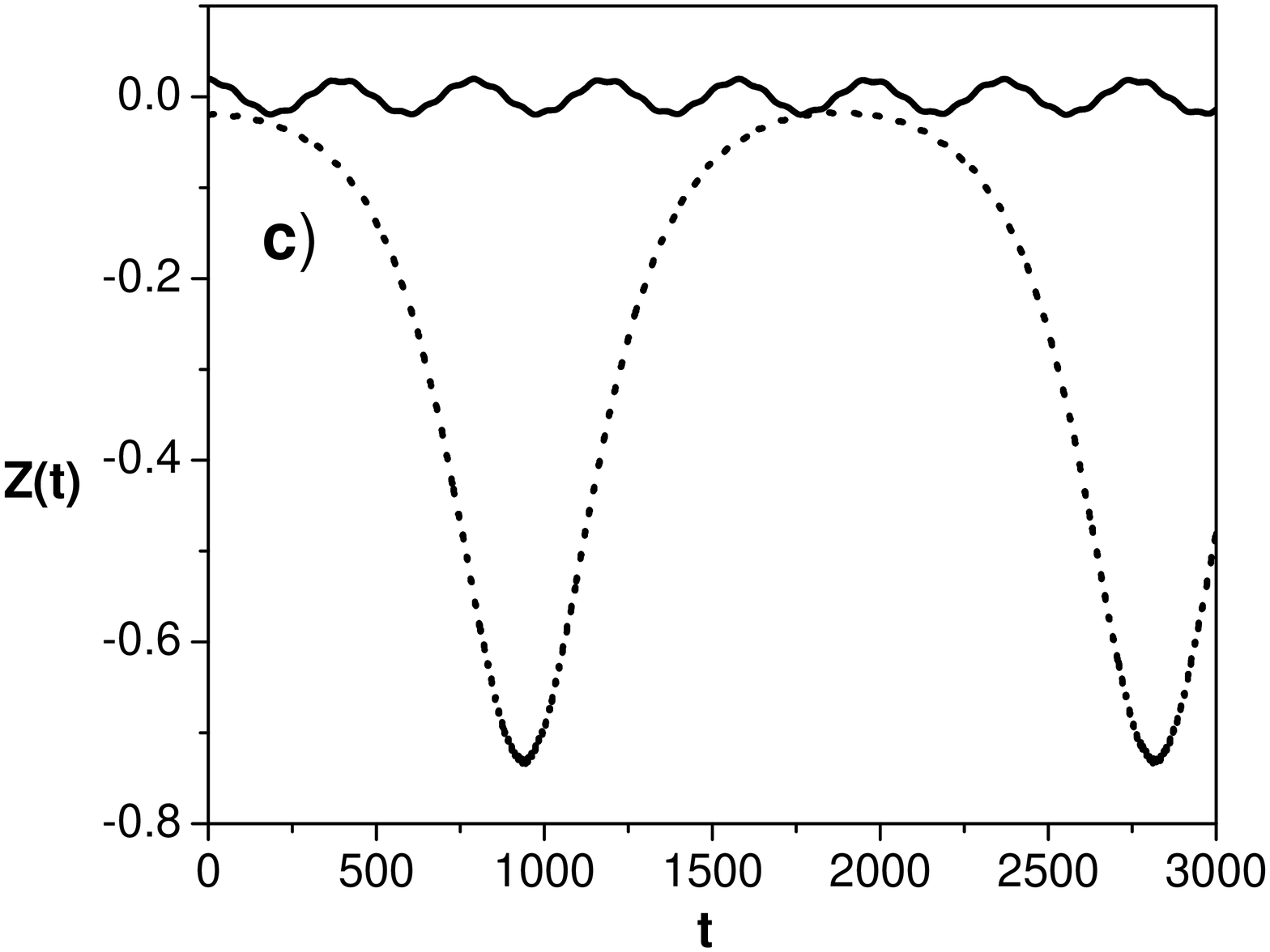}\quad
\caption{The time dependencies of $Z(t)$ for the LHY-fluid (with
$\delta g =0$): a) $\gamma = 0.02 \ $, b) $\gamma = 0.025 \ $, c)
$\gamma = 0.03$. Solid lines are for the zero-phase and dot-lines
are for the $\pi$-phase modes. The parameters of the trap
potential are $V_0=7, \ l=1.5$. Everywhere initial imbalance
$Z(t=0)=0.02$.} \label{bifurcation}
\end{figure}
It is interesting to examine the evolution of the time
dependencies of the population imbalance $Z(t)$ of the LHY-fluid
near the symmetry breaking point that one can see at the beginning of the
lower curve depicted in Fig.~\ref{LHYdg}. In Fig.~\ref{bifurcation}
the time dependencies of $Z(t)$ for three values of $\gamma$ are
presented. One can see a transition of the $\pi$-phase mode into near
the bifurcation point located between $\gamma=0.2$ and
$\gamma=0.25$.

We consider  Josephson oscillations of the atomic population
imbalance $Z(t)$ ($\langle Z \rangle =0$) in the quasi-linear and
self-localization regimes ($<Z> \neq 0$).

\begin{figure}[htbp]
\includegraphics[width=5.0cm,clip]{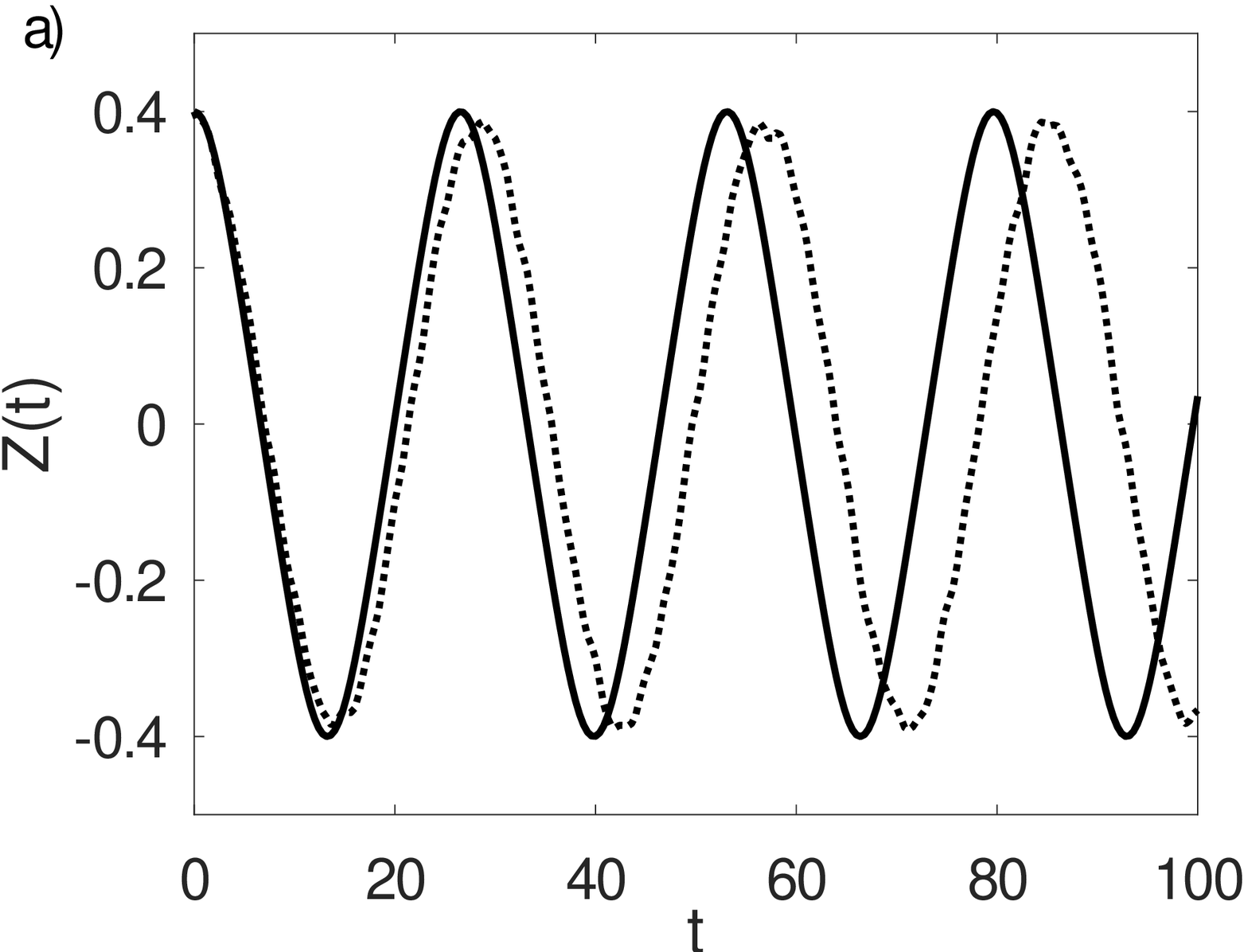}\quad
\includegraphics[width=5.0cm,clip]{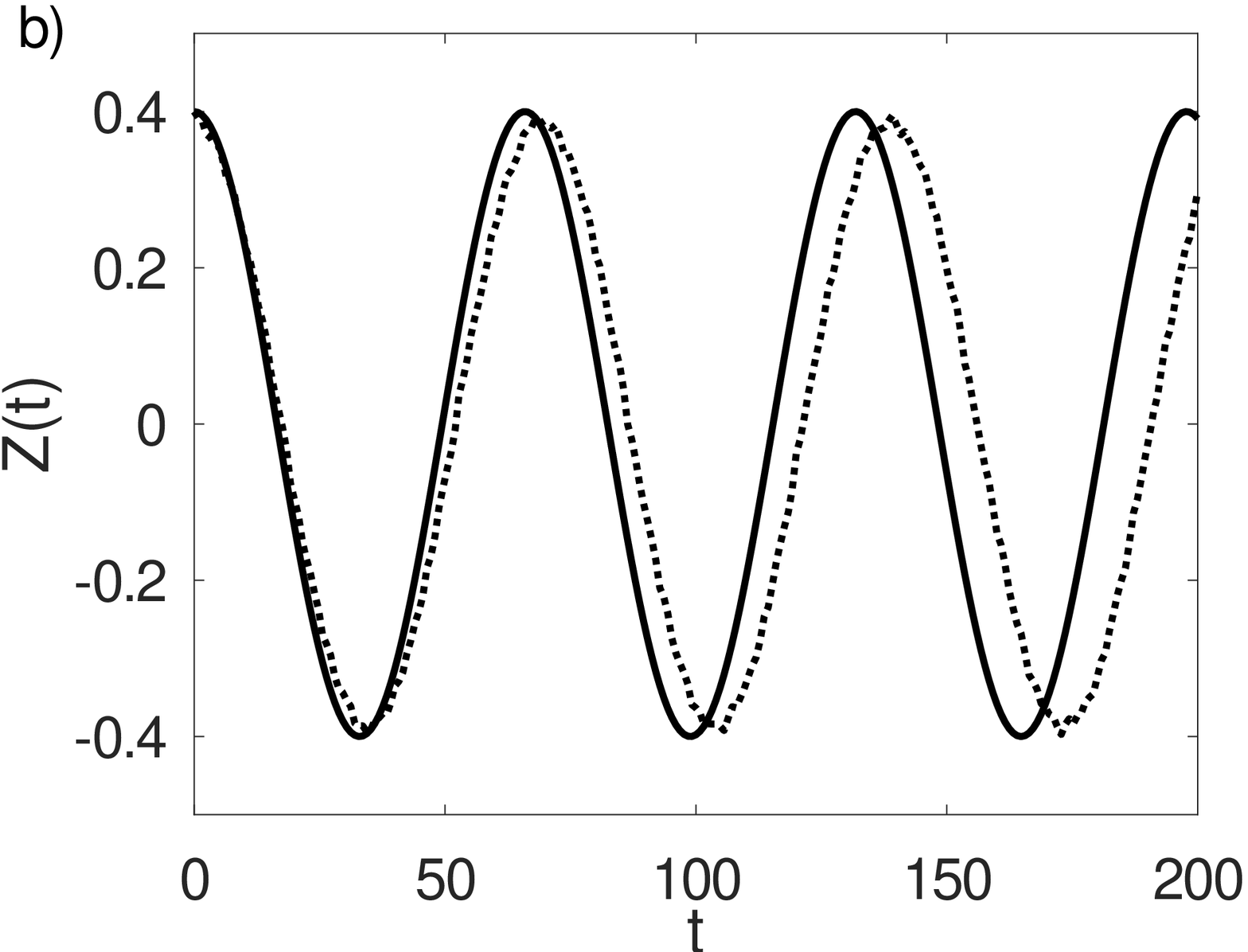}\quad
\includegraphics[width=5.0cm,clip]{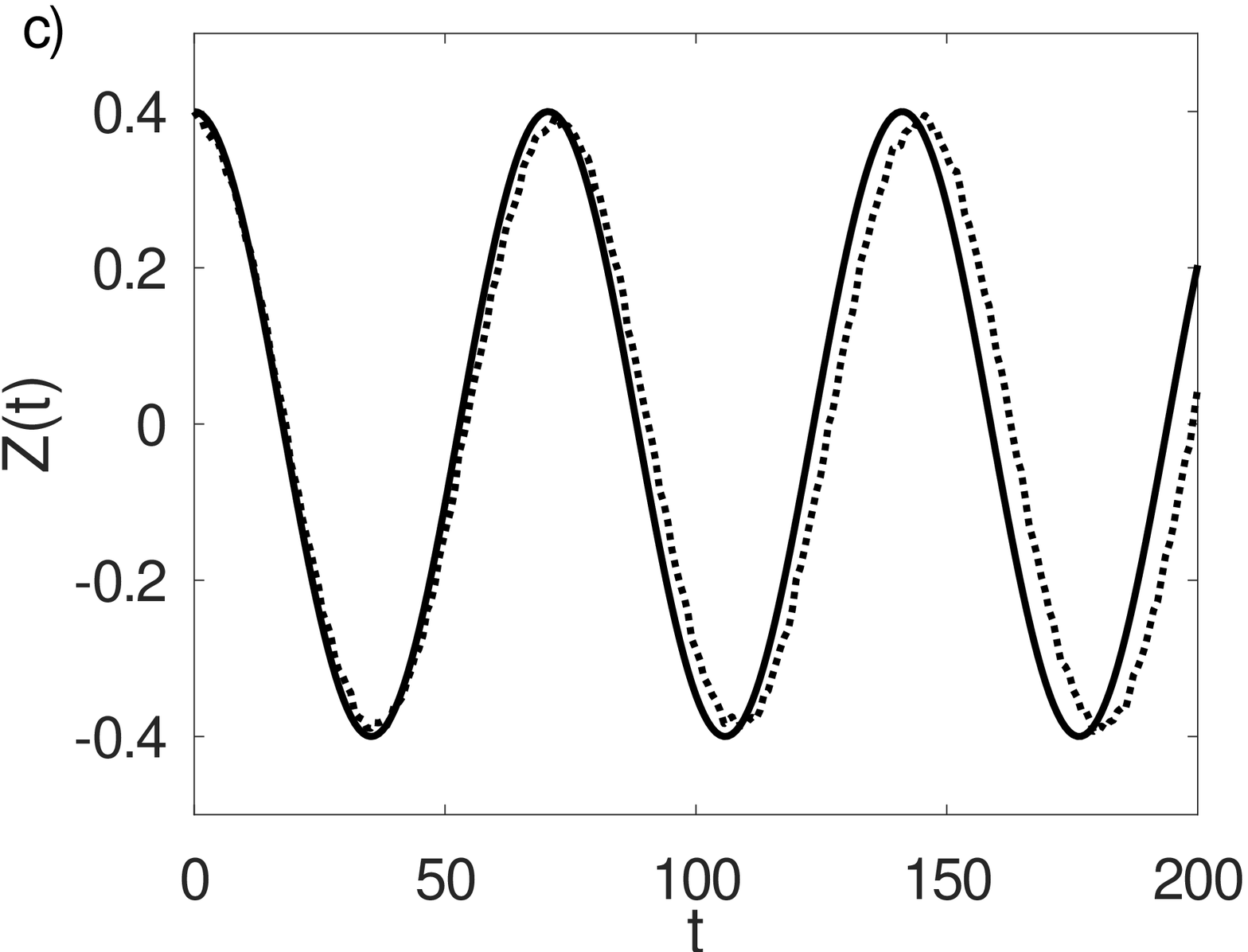}\quad
\caption{The time dependencies of the imbalance $Z(t)$ obtained
from the two-mode model (solid lines) and the full numerical
solution of Eq~(\ref{eq1D}) (dot lines) for: a) $\delta g = 1$,
$\gamma =0$, (zero-phase); b) $\delta g = 0$, $\gamma=0.01$,
(zero-phase); c) $\delta g = 0$, $\gamma=0.01$ ($\pi$-phase).
Everywhere initial imbalance $Z(0)=0.4$. The parameters of the
trap potential are $V_0=4, \ l=1.5$.} \label{V04JO}
\end{figure}

In Fig.~\ref{V04JO}, Josephson oscillations are presented
based on the results obtained from the two-modes model and the
numerical simulations. In all graphs, one can see phase shifts
that increase with time.

In Fig.\ref{V04JO}a, the largest phase shift is observed when the
nonlinear parameter $\delta g=1$ is large enough (at $\gamma=0$).
For the LHY-fluid case, in Figs.~\ref{V04JO}b and \ref{V04JO}c
with the nonlinear parameters $\delta g=0$ and $\gamma=0.01$
(small enough), the phase shifts decrease. So, the accuracy of the
two-modes method decreases at large times and large values of
nonlinear parameters.


\begin{figure}[htbp]
\includegraphics[width=5.0cm,clip]{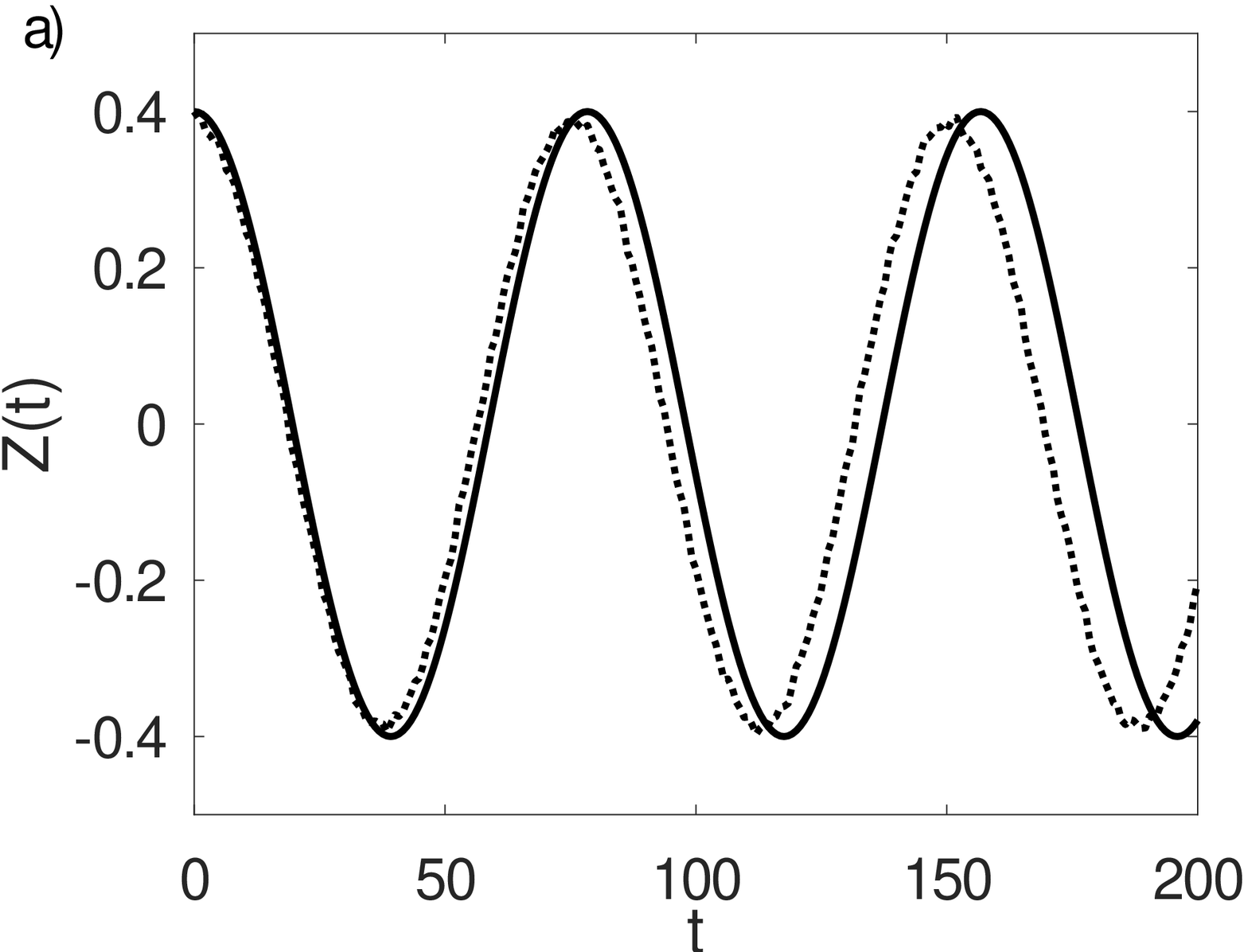}\quad
\includegraphics[width=5.0cm,clip]{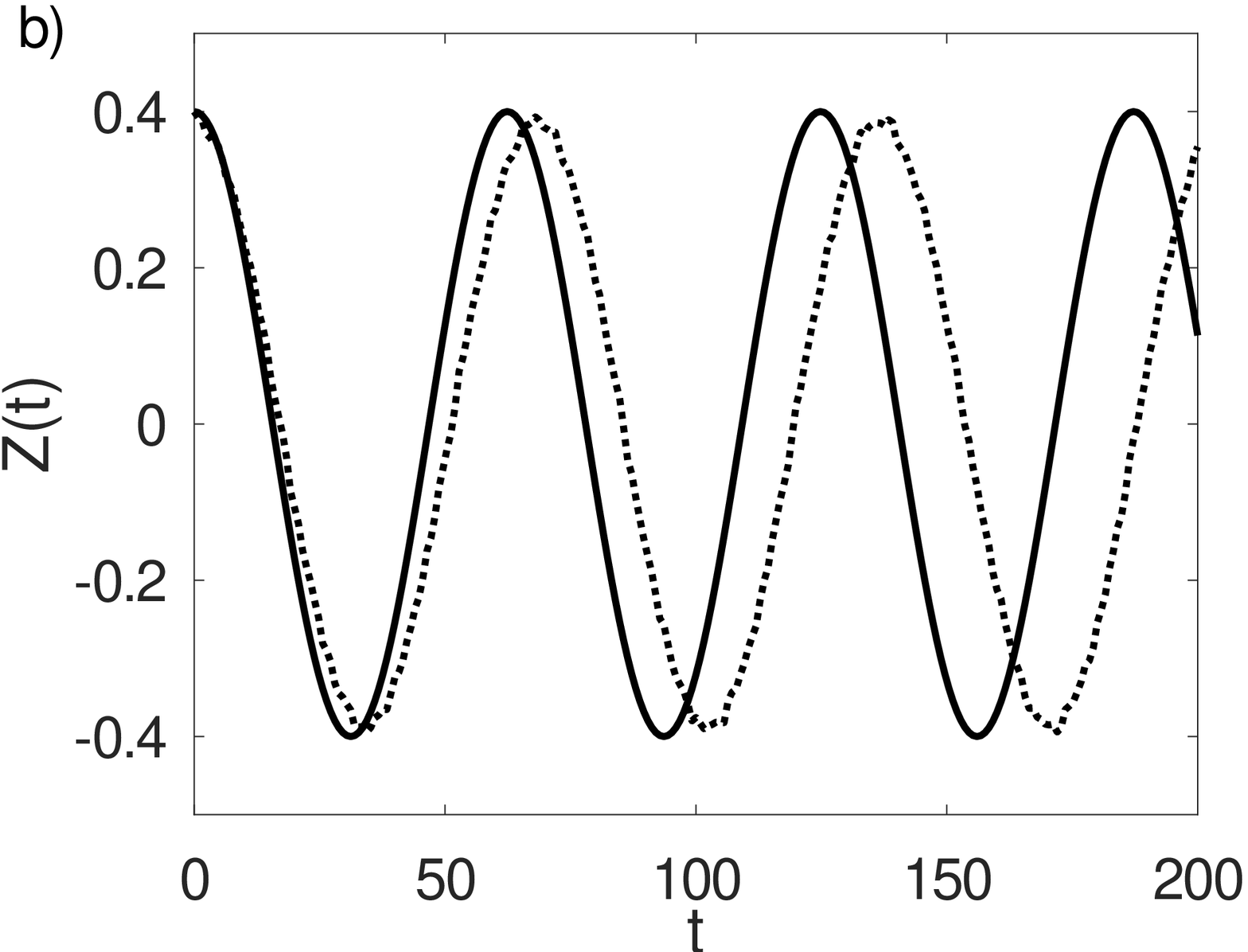}\quad
\includegraphics[width=5.0cm,clip]{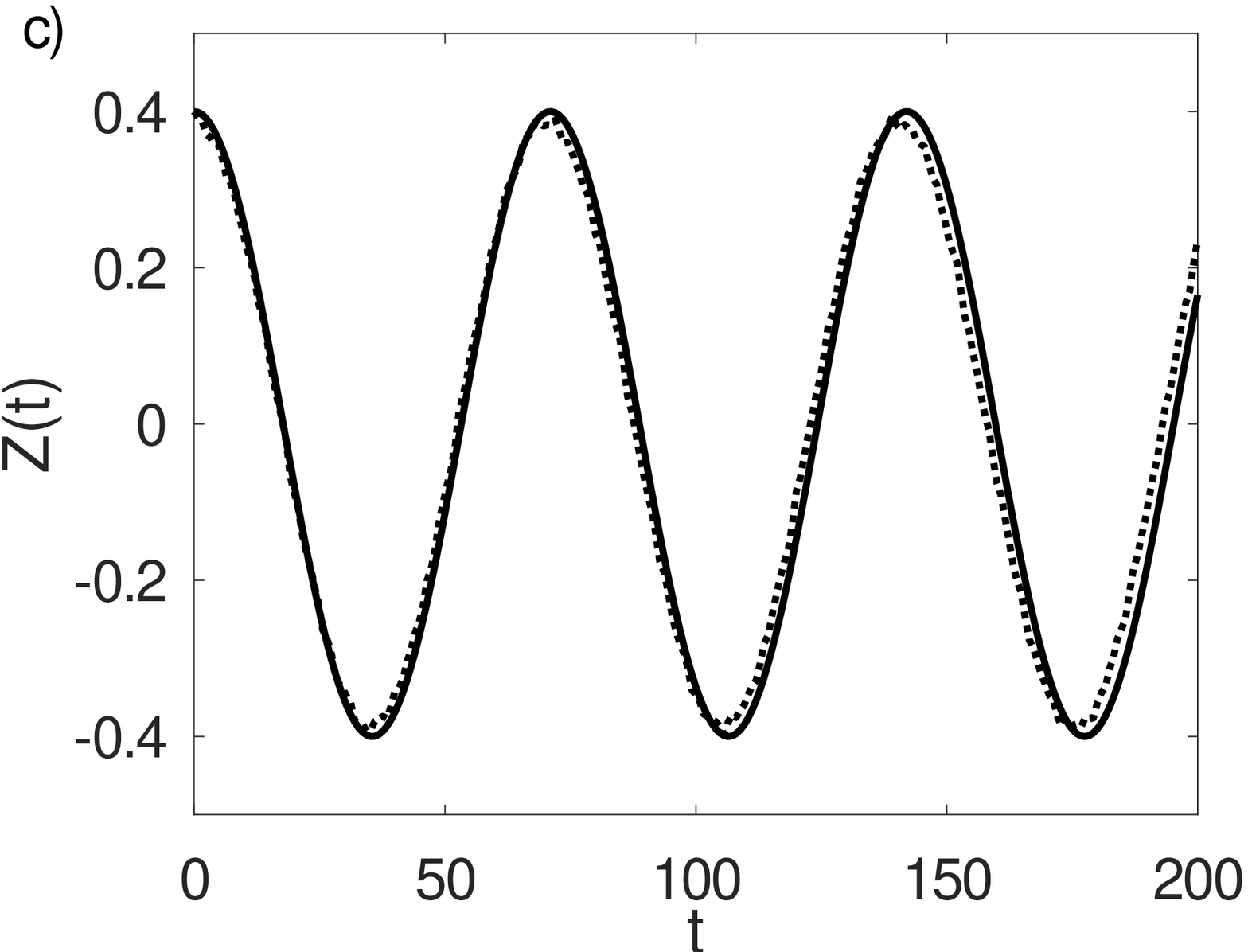}\quad
\caption{The time dependencies of the imbalance $Z(t)$ obtained
from the two-modes model (solid lines) and the full numerical
solution of Eq~(\ref{eq1D}) (dot lines) for: a) $\delta g = -0.1$,
$\gamma =0.1$ (zero-phase); b) $\delta g =-0.1$, $\gamma=0.1$
($\pi$-phase), c) $\delta \ g =-0.076$, $\gamma=0.1$ (zero-phase)
with initial imbalance $Z(0)=0.4$. The parameters of the trap
potential are $V_0=4, \ l=1.5$.} \label{V04QL}
\end{figure}

From Eq.~(\ref{eq1D}) we see that if two nonlinear parameters are
taken in opposite signs, they can suppress each other and
\textbf{the quasilinear} (the Rabi) regime can be obtained. To create this
regime we should calculate necessary values of the nonlinear
parameters using Eq.~(\ref{quasilinear}). Thus  we need also to
calculate parameters $\alpha = \int_{-\infty}^{\infty}
|\Phi_1(x)|^4 dx$ and $\beta = \int_{-\infty}^{\infty}
|\Phi_1(x)|^5 dx$ that depend on values of the nonlinear
parameters $\delta g$ and $\gamma$. In Figs.~\ref{V04QL}a and
\ref{V04QL}b, in order to observe effect of suppression, we choose
the nonlinear parameters with opposite signs, $\delta g=-0.1, \
\gamma=0.1$ at the zero-phase and the $\pi$-phase modes.
In Fig.~\ref{V04QL}c we choose corrected
value for $\delta g$:  $\delta g=-0.076, \ \gamma=0.1$ at the
zero-phase mode. calculated from Eq.(\ref{quasilinear}). One can
see from Figs.~\ref{V04QL}a and \ref{V04QL}b that there remain the
phase shift between the solutions calculated based on the
two-mode model and numerical simulations of the GP equation.
Fig.~\ref{V04QL}c shows that the phase shifts of solutions
obtained on the basis of the two-mode method and the numerical
method have almost disappeared which indicates that the nonlinear
effects have decreased and the mode of quasi-linear oscillations  has been reached.

\begin{figure}[htbp]
\includegraphics[width=4.0cm,angle=-90,clip]{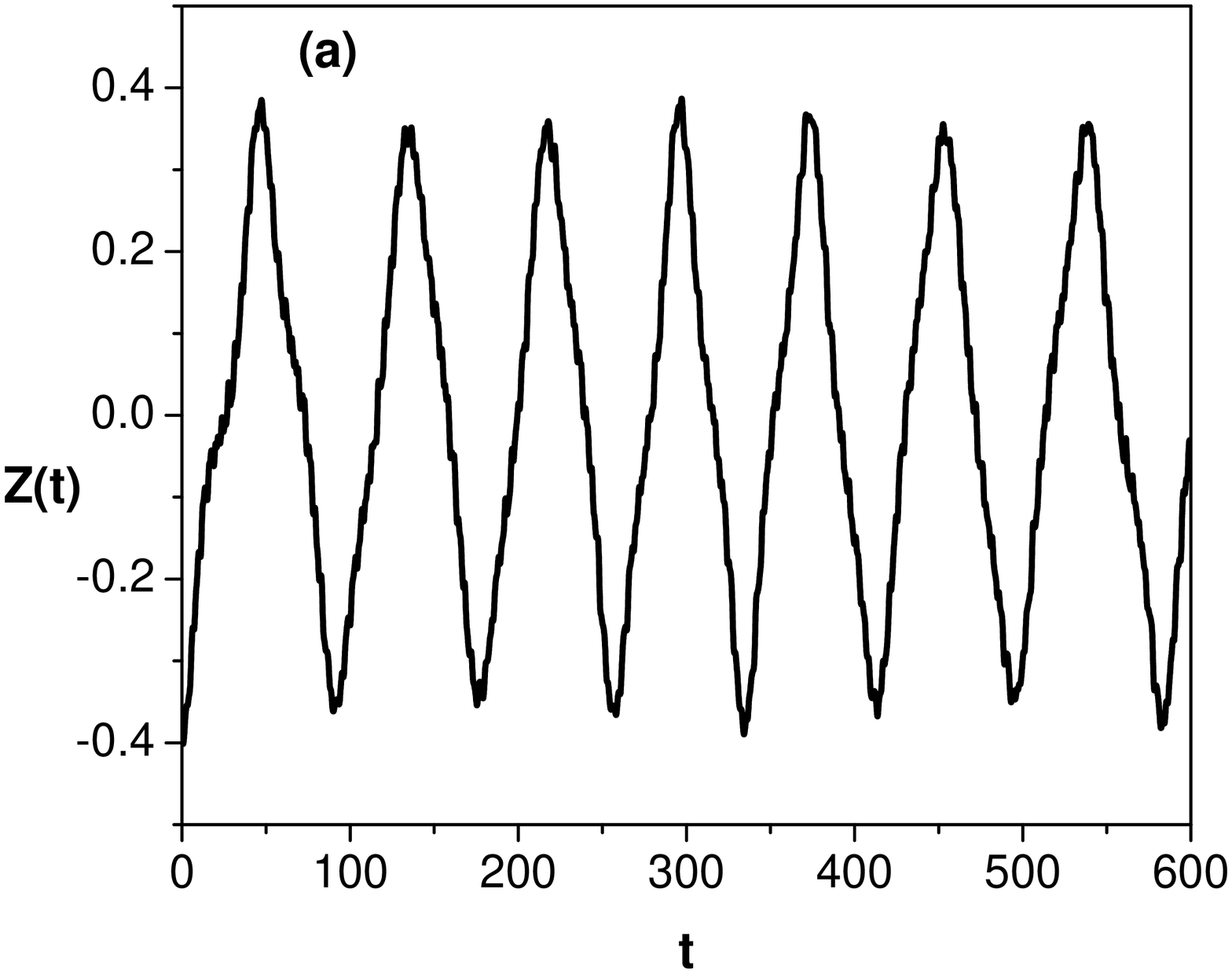}\quad
\includegraphics[width=4.0cm,angle=-90,clip]{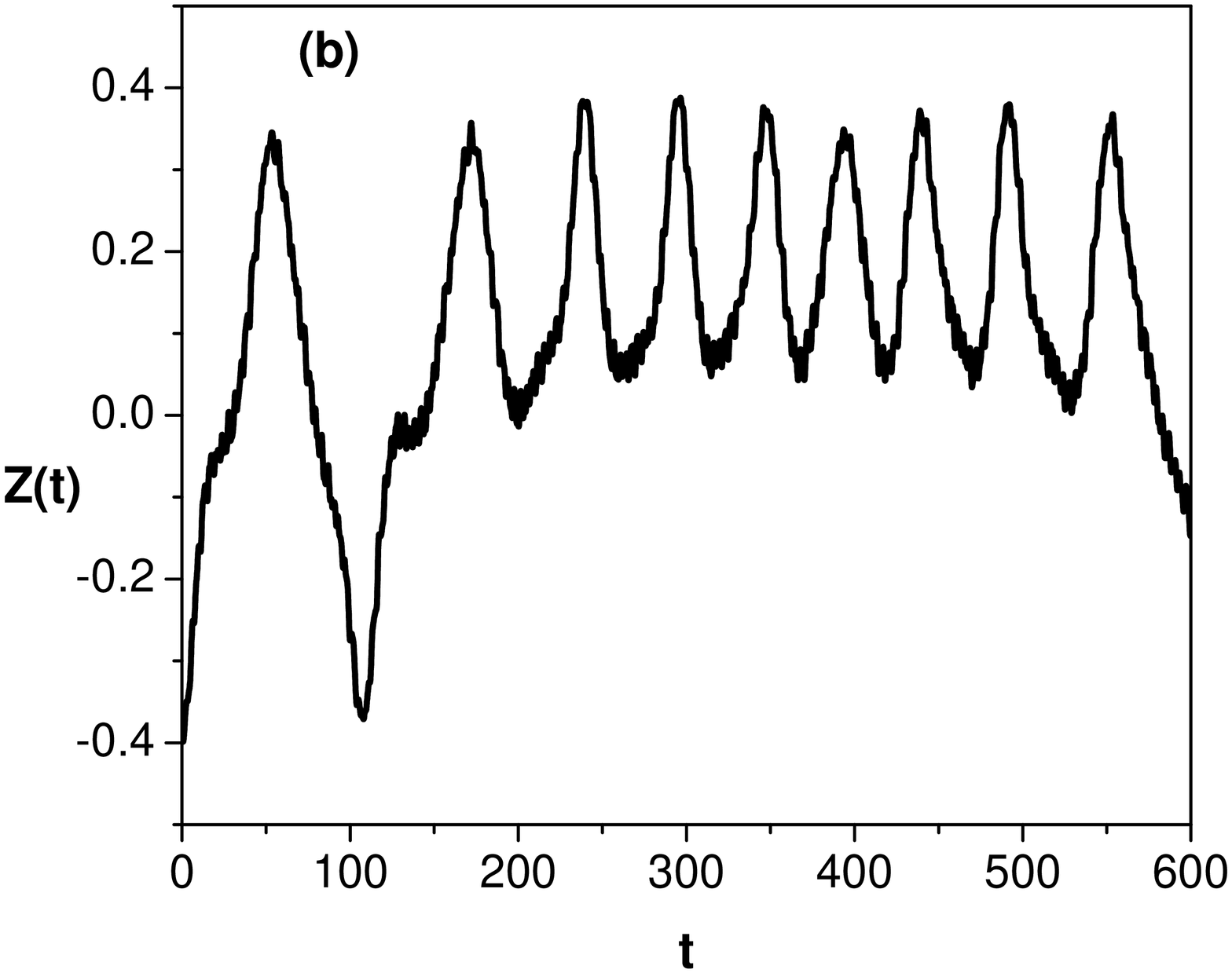}\quad
\includegraphics[width=4.0cm,angle=-90,clip]{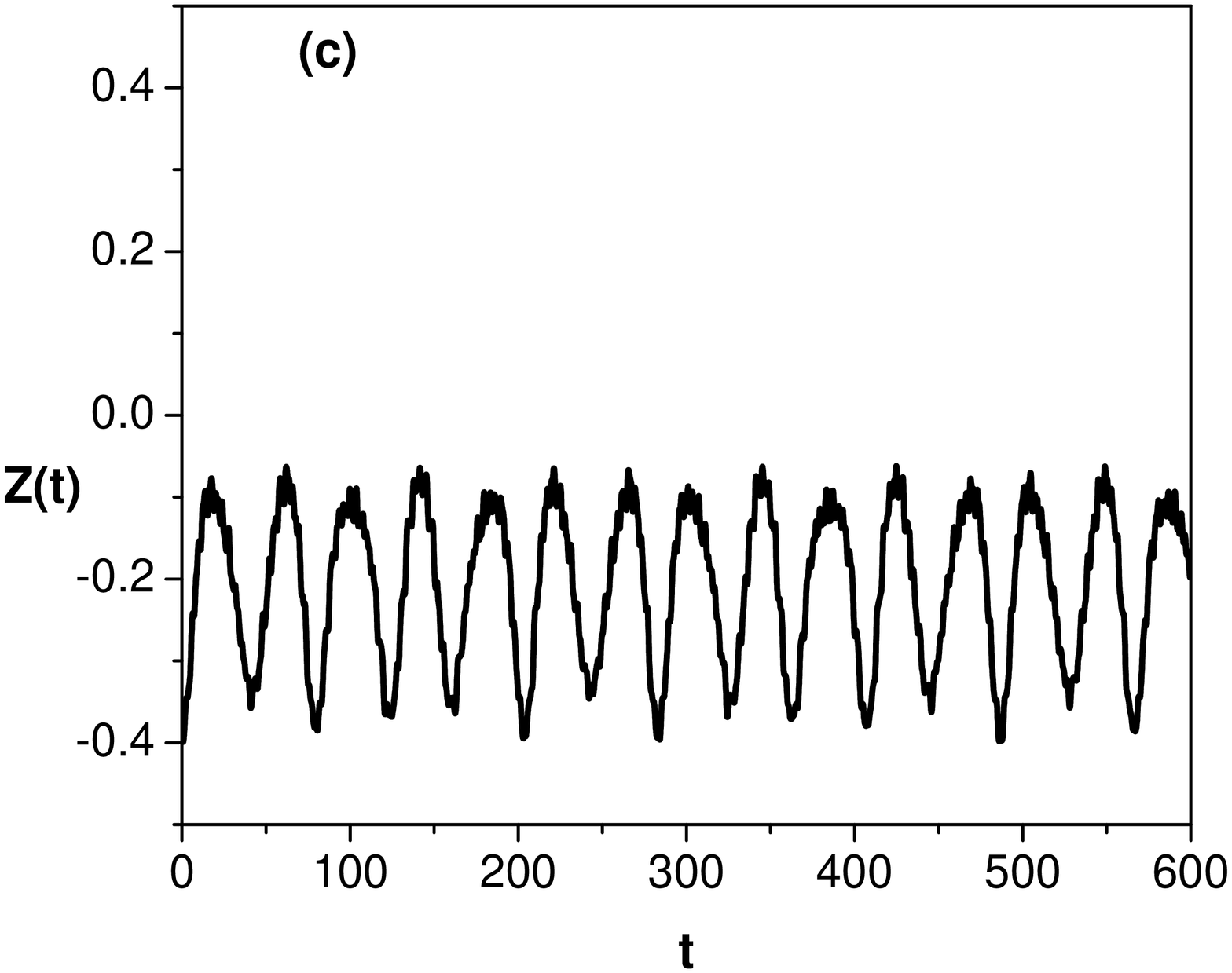}\quad
\caption{The time dependencies of $Z(t)$ for the LHY-fluid case
(with $\delta g=0$) when the LHY-term is  a) $\gamma=1.5$; b)
$\gamma=1.6$; c) $\gamma=1.8$. The parameters of the trap
potential $V_0=6, \ l=1.5$.} \label{V06LHY}
\end{figure}

To study the localization phenomenon, we consider the case of the LHY-fluid,
when $\gamma \neq 0$ at $\delta g=0$. The results of calculation
are given in Fig.~\ref{V06LHY} for three values of $\gamma= 1.5; \
1.6; \ 1.8$. As seen, the self-localization phenomenon is observed
at $\gamma \simeq 1.6$. Two-mode model predict the existence of the ST transition, but the agreement
with full numerical simulations of the MGP equation is only qualitative. This discrepancy  is connected with neglect of
nonlinear overlap integrals and thus to the error in the calculation of the tunneling amplitude $K$ \cite{AB06}.


\begin{figure}[htbp]
\includegraphics[width=4cm,angle=-90,clip]{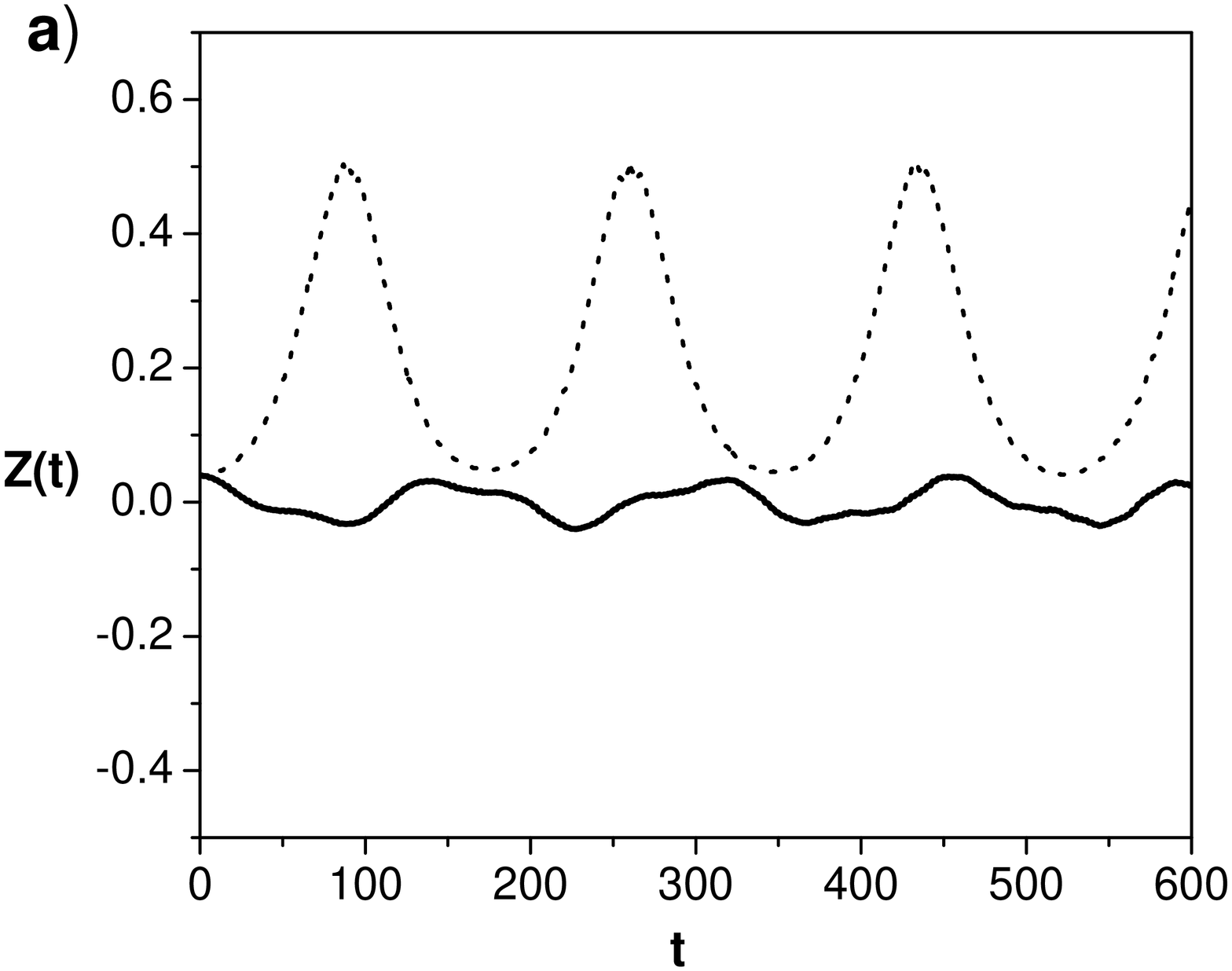}
\includegraphics[width=4cm,angle=-90,clip]{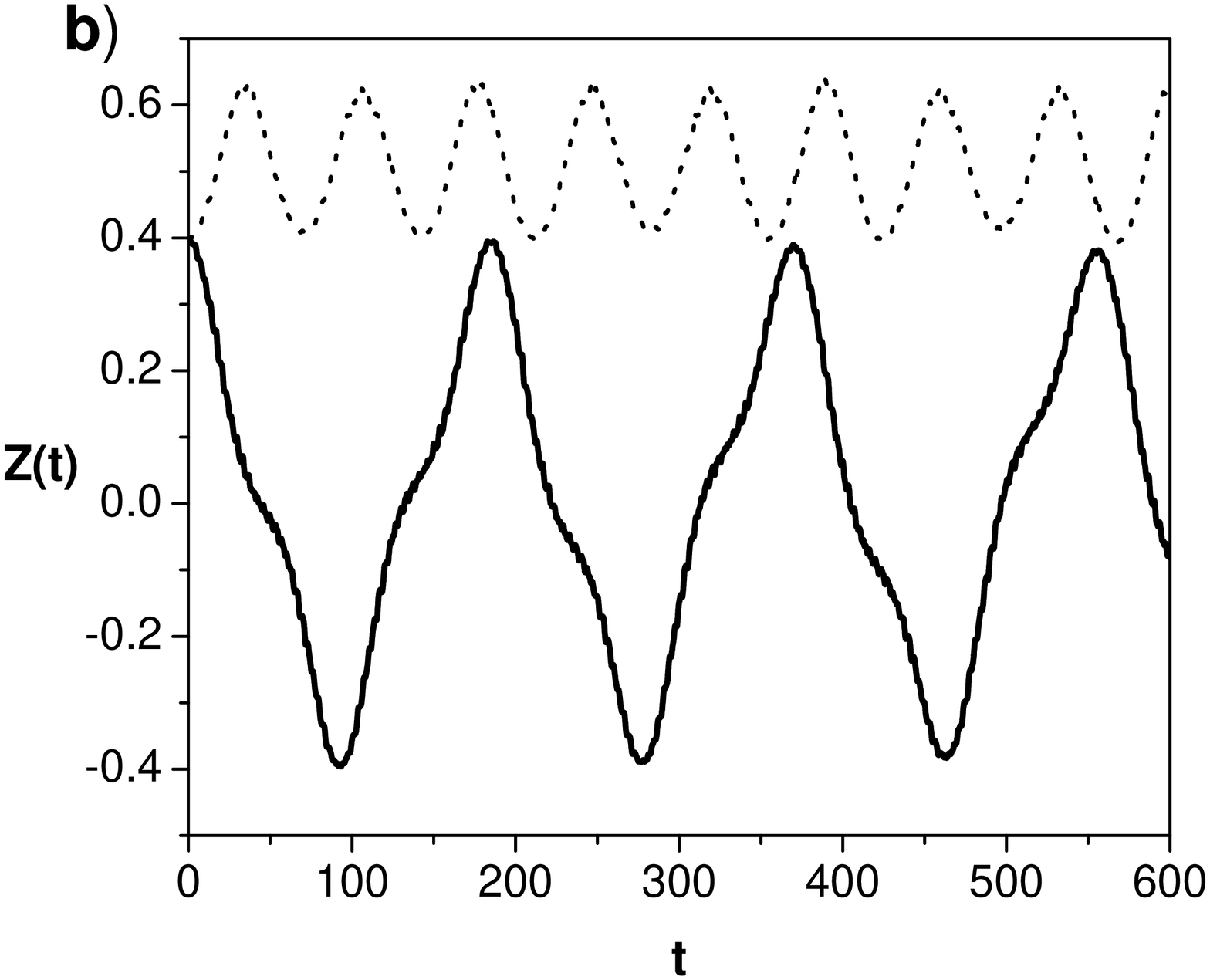}
\includegraphics[width=4cm,angle=-90,clip]{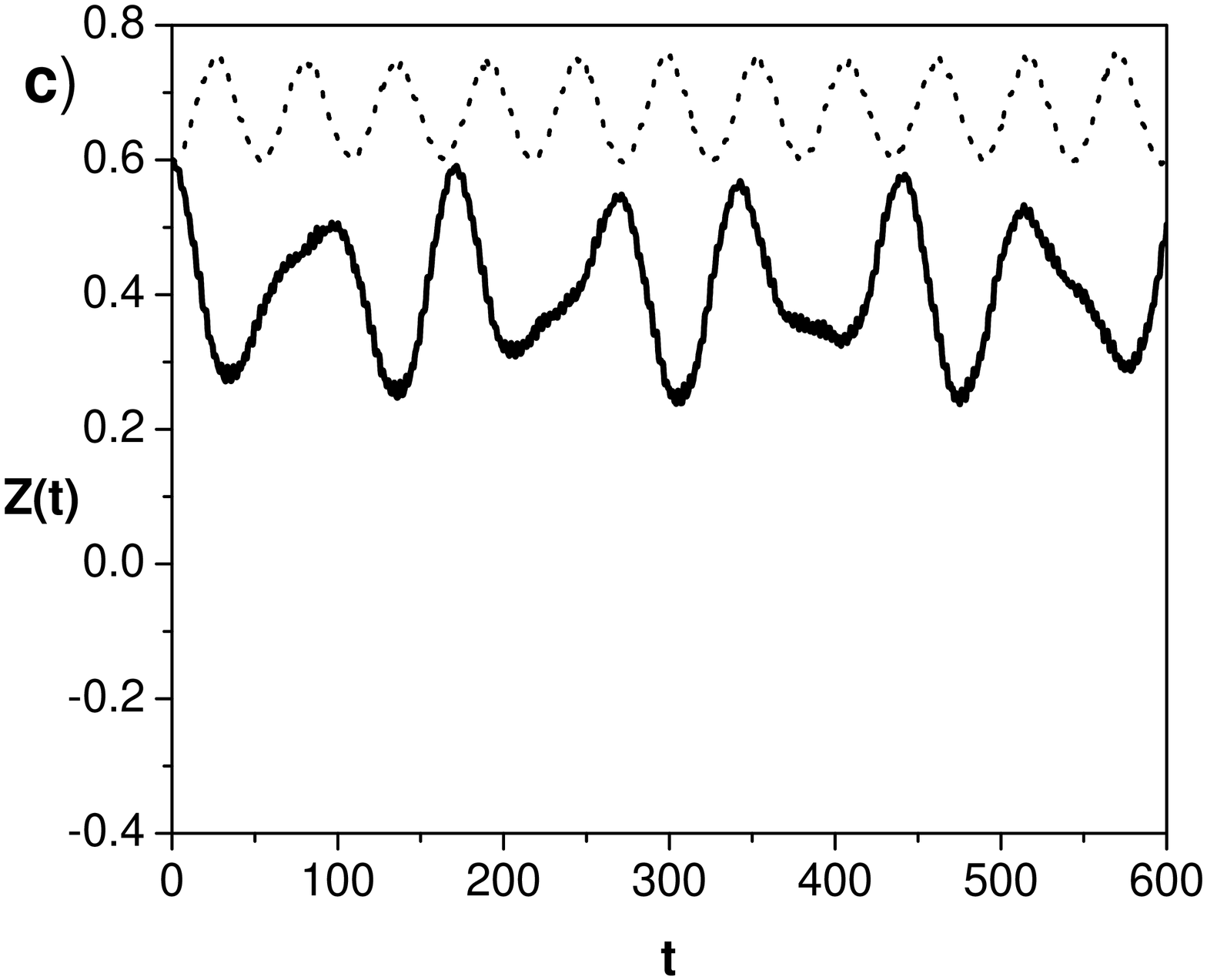}
\caption{Time dependencies of $Z(t)$ for different initial
imbalances: a) $Z(0)=0.4$; b) $Z(0)=0.6$; c) $Z(0)=0.6$. All
graphs are presented for $\delta g=-0.1$, $\gamma=0.5$. Solid
lines are for the $\pi$-phase $\psi(0)=\pi$ and dot lines are for
the zero-phase $\psi(0)=0$ modes. Parameters of the trap potential
$V_0=7, \ l_0=1.5$.} \label{dggLHY}
\end{figure}

In Figs.~\ref{dggLHY}a, \ref{dggLHY}b and \ref{dggLHY}c the time
dependencies of the imbalance $Z(t)$ at zero-phase and $\pi$-phase
modes for $\gamma=0.5$, $\delta g=-0.1$ are presented for
different values of the initial imbalance. As seen, an increase in
the initial imbalance $Z(t=0)$ causes a change in the behavior of
the dependencies $Z(t)$ and the emergence of the localization in
both phase modes.

Let us estimate the parameters for the experiment with $^{26}$Ka BEC, as performed in Ref.~\cite{Skov}. Transverse and longitudinal frequencies of the trap can be taken as, $\omega_{\perp} =2\pi \cdot 200$Hz, $\omega_{x}=2\pi \cdot 40$Hz and the number of atoms is $N= 10^3 \div 10^4$.
Double-well potential can be generated by the superposition of the harmonic trap and periodic potential\cite{expDW}. The barrier values for this double-well potential are $2\pi \cdot (160-240)$Hz when $V_0=4-6$.
 The atomic scattering length is about $a=50a_0$ and the scattering length between two components is about $\delta a=a_0$,
 where $a_0$ is the Bohr radius. The values of the scattering length  can be varied  using the Feshbach resonance technics. If we take $a=57.5a_0$, $\delta a=0$ and $N=10^4$, we get the values $\gamma=1.5$ and $\delta g=0$ and JO frequencies is $\omega_J=2\pi \cdot 2.8$Hz (Josephson oscillation around $\langle Z \rangle =0$). If we take $a=61.9a_0$, $\delta a=0$ and $N=10^4$, we get the values $\gamma=1.8$ and $\delta g=0$ and frequencies of atomic population imbalance $Z(t)$ is $\omega_J =2\pi \cdot 6$Hz(the frequency of self-localization mode) (see~Fig.\ref{V06LHY}). For the case  $a=148a_0$, $\delta a=-0.25a_0$ and $N=10^3$, we get the values $\gamma=0.5$ and $\delta g=-0.1$ and thus frequencies of atomic population imbalance $Z(t)$ is  $\omega_J=2\pi \cdot 1.35$Hz for zero-phase mode (Josephson oscillation around $\langle Z \rangle =0$) and $\omega_J=2\pi \cdot 3.6$Hz for $\pi$-phase mode(self-localization mode)  (see Fig.\ref{dggLHY}a).

\section{Conclusion.}
\label{sec:conclusion}

In conclusion, we have investigated the influence of  quantum
fluctuations on the dynamics of a two-component BEC in the double-well potential.  In the scalar
approximation and in a quasi-one-dimensional geometry, the problem
is described by the modified GP equation with the additional LHY
terms, corresponding to the beyond mean-field effects of quantum
fluctuations.

We have derived a two-mode (dimer) model to describe the macroscopic
quantum tunneling and self-trapping phenomena, when the quantum
fluctuations are taken into account. The frequencies of the
Josephson oscillations for zero- and $\pi$- phase modes are
calculated. It is shown that the quasi-linear (i.e. the Rabi oscillations) regime of
oscillations can be observed by a proper choice of the residual
mean-field interactions between components of the  BEC. It opens the
possibility to measure the  strength of quantum fluctuations.

The criterion for switching from the MQT regime to the ST one is
obtained and applications for the observation of the LHY fluid have been
discussed.  Direct simulations of the modified GP equation with
the Lee-Huang-Yang term, confirm predictions of the two-mode
model for the JO frequencies. For the self-trapping regime, the
agreement is only qualitative, due to limits of the two-mode
model. Also, full simulations predict the self-trapping regime of
the LHY fluid in the double-well potential.

In future will be interested to investigate the MQT and MST phenomena in the crossover regime from 3D and 2D
geometry to one-dimensional geometry.

\section{Acknowledgments}
We thank B.Baizakov and  E. N. Tsoy for fruitful discussions. The
work was supported by the grant FA-F2-004 of Ministry of
Innovative Development of the Republic of Uzbekistan.
 

\vspace{1cm}
\begin{thebibliography}{99}

\bibitem{LHY}

Lee T D, Huang K, and Yang C N
 1957 \emph{Phys. Rev.} {\bf 106} 1135.


\bibitem{Petrov}

Petrov D S 2015 \emph{Phys.Rev.Lett.} {\bf 115} 155302.

\bibitem{PA}

Petrov D S and Astrakharchik G E 2016 \emph{Rev. Lett.} {\bf 117} 100401.

\bibitem{expmixt}
Cabrera C R, Tanzi L, Sanz J, Naylor B, Thomas P, Cheiney P, and Tarruell L 2018 \emph{Science} {\bf 359} 301;
Semeghini G, Ferioli G, Masi L, Mazzinghi C, Wolswijk L, Minardi F, Modugno M, Modugno G, Inguscio M, and Fattori M 2018 \emph{Phys. Rev. Lett.} {\bf 120} 235301;
D\'Errico C,  Burchianti A, Prevedelli M, Salasnich L, Ancilotto F, Modugno M, Minardi F, and Fort C 2019 \emph{Phys. Rev. research} {\bf 1} 033155

\bibitem{Dip1}
Ferrier-Barbut I, Kadau H, Schmitt M, Wenzel M, Pfau T
2016 \emph{Phys. Rev. Lett.} {\bf 116} 215301.

\bibitem{BF}
Rakshit D, Karpiuk T, Brewczyk M and Gajda M 2019 \emph{SciPost Phys.} {\bf 6} 079

\bibitem{Rabi}
Cappellaro A, Macr T,  Bertacco G  F, Salasnich L 2017
\emph{Scientific reports} {\bf 7(1)} 13358

\bibitem{solit}
Astrakharchik G E and Malomed B A, 2018 \emph{Phys. Rev. A} {\bf 98} 013631;
Abdullaev F Kh, Gammal A, Kumar R K and Tomio L 2019 \emph{J. Phys. B: At. Mol. Opt. Phys.} {\bf 52} 195301;
Otajonov Sh R, Tsoy E N, and Abdullaev F Kh 2019 \emph{Physics Letters A} {\bf 383}  125980.

\bibitem{collexc}
Pathak M  R, Nath A 2022 \emph{Scientific Reports} {\bf 12} 6904

\bibitem{MI}
Otajonov Sh R, Tsoy E N, and Abdullaev F Kh 2022 \emph{Phys. Rev. A} {\bf 106} 033309;  Kartashov Y  V, Lashkin M, Modugno M, and Torner L 2022 \emph{New J. Phys.} {\bf 24}  073012.

\bibitem{Jor}
Jorgensen N B et al. 2018 \emph{Phys.Rev. Lett.} {\bf 121} 173403

\bibitem{TM}
Raghavan S, Smerzi A, Fantoni  S, and Shenoy S R 1999 \emph{Phys. Rev. A} {\bf 59} 620

 \bibitem{Skov}
Skov T G et al. 2021 \emph{Phys.Rev.Lett.} {\bf 126} 230404

\bibitem{Michinel}
P\'erez-García M V, Michinel H, Cirac J I, Lewenstein M, and Zoller P
1997 \emph{Phys. Rev. A} {\bf 56} 1424

\bibitem{AB06}
Ananikian D and Bergeman T 2006 \emph{Phys. Rev. A} {\bf 73} 013604

\bibitem{Bao}
Xiong B, Gong J, Han Pu, Bao W, and Li B 2009 \emph{Phys. Rev. A} {\bf 79} 013626

\bibitem{expDW}
Albiez M, Gati R, Fölling J, Hunsmann S, Cristiani M, and Oberthaler M K 2005
\emph{Phys. Rev. Lett.} {\bf 95} 010402

\end{thebibliography}
\end{document}